# FIRI
## A Far-InfraRed Interferometer for ESA

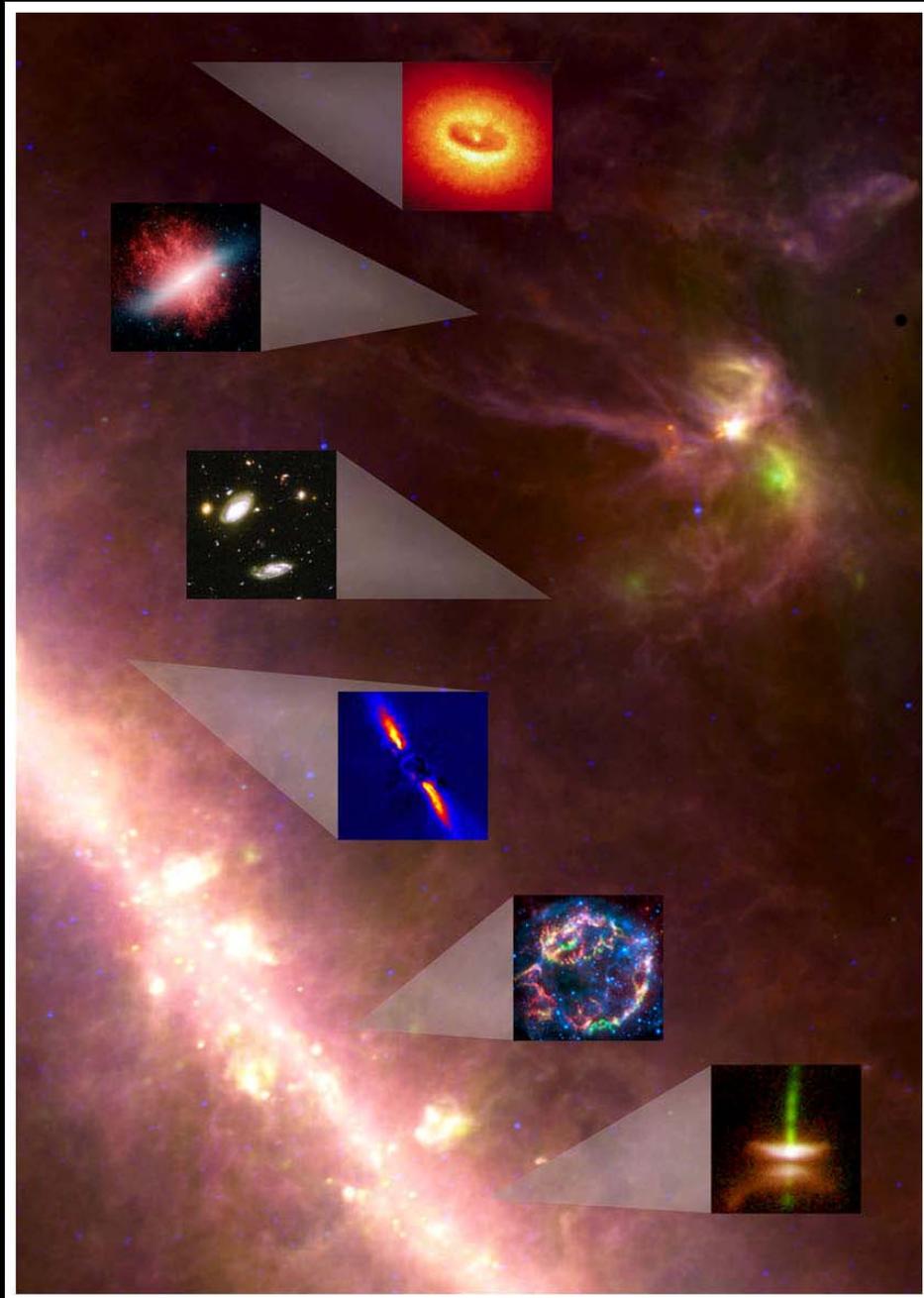

## Proposal for Cosmic Vision 2015-2025

Frank Helmich & Rob Ivison

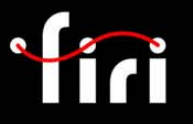

# FIRI – a Far-InfraRed Interferometer for ESA
Dr Frank Helmich, SRON, NL & Prof. Rob Ivison, UK ATC

## *Index*

## *Executive Summary*

Half of the energy ever emitted by stars and accreting objects comes to us in the far-infrared (FIR) waveband and has yet to be properly explored. We propose a powerful Far-InfraRed Interferometer mission, *FIRI*, to carry out high-resolution imaging spectroscopy in the FIR. This key observational capability is essential to reveal how gas and dust evolve into stars and planets, how the first luminous objects in the Universe ignited, how galaxies formed, and when super-massive black holes grew. *FIRI* will disentangle the cosmic histories of star formation and accretion onto black holes and will trace the assembly and evolution of quiescent galaxies like our Milky Way. Perhaps most importantly, *FIRI* will observe all stages of planetary system formation and recognise Earth-like planets that may harbour life, via its ability to image the dust structures in planetary systems. It will thus address directly questions fundamental to our understanding of how the Universe has developed and evolved – the very questions posed by ESA's Cosmic Vision:

- What are the conditions for stars to form, and where do they form?
- How do stars evolve as a function of their interstellar environment?
- In which conditions do planets form around stars?
- How were the first luminous objects in the Universe ignited? How did the first stars form and evolve?
- How did the history of stars and supernovae give rise to current chemical element abundances?
- What is the history of super-massive black holes and how do they interact with their host galaxy?
- What is the nature of the FIR background, and of early, deeply embedded star formation?

The FIR region of the electro-magnetic spectrum is the last major band where poor angular resolution and lack of sensitivity hinders progress. Pathfinders *Herschel* and *SPICA* will provide major advances in sensitivity, but will lack the angular resolution necessary to resolve the cosmic FIR background radiation or to undertake detailed studies of individual objects. ALMA and the *James Webb Space Telecope (JWST)* will provide high angular resolution and sensitivity at shorter and longer wavelengths, but the crucial band between 25 and 300 µm is not covered by any comparable instrument: there exists a crippling lack of observational capability in the FIR, despite the vital role this band plays in exploring the formation and evolution of active galactic nuclei (AGN), galaxies, stars and planetary systems, and the development of life-sustaining environments.

This "FIR gap" (Figure 1) is recognised by the astrophysical community and has been noted by ESA's Astronomy Working Group. In the ESLAB 2005 Cosmic Vision symposium, a high-angular-resolution FIR observatory with high angular resolution was listed as a major priority for ESA's science programme: *FIRI* is such a mission, strongly supported by the worldwide astronomy community and already studied extensively by ESA, NASA and others.

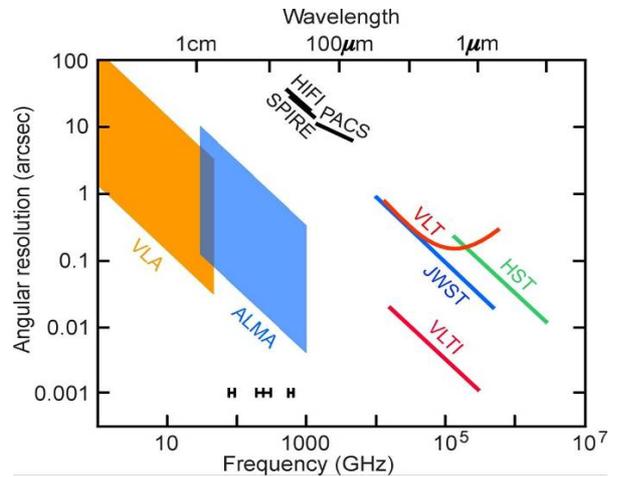

**Figure 1:** With the advent of ALMA and Very Large Telescope Interferometry (VLTI), the FIR gap is deepening.

*FIRI* will make unique and key contributions to our understanding of the Universe, near and far. It will peer through the dust that shrouds stellar nurseries in our Galaxy, de-mystifying the process by which stars and planets are born. It will image proto-stellar and debris disks at the peak of their spectral energy distributions (SEDs), where the brightness is 1000× that at a wavelength of 1 mm, with exquisite spectral resolution and sensitivity, revealing how planetary systems form out of gas, dust and ice. While the *Hubble Space Telescope (HST)* has produced beautiful pictures of merging and star-forming galaxies, crude submillimetre observations have shown that the real action is in the FIR. *FIRI*'s angular resolution will break through the confusion limit and allow us to determine the properties and internal structure of distant star-forming galaxies, and to examine the enigmatic symbiosis between host galaxies and their AGN. The earliest metal-poor galaxies will be very highly redshifted (i.e. at large look-back times), and we may even be able to detect their formation via molecular hydrogen emission red-shifted into the FIR. This would provide a unique probe of first light - the formation of the earliest stars in the Universe and the ensuing re-ionisation of the Universe.

Here, we outline the *FIRI* mission concept – three cold, 3.5-m apertures, orbiting a beam-combining module, with separation of up to 1 km, free-flying or tethered, operating between 25 and 385 µm, using the interferometric direct-detection technique to ensure µJy sensitivity and 0.02" resolution at 100 µm, across an arcmin$^2$ instantaneous field of view, with a spectral resolution, $\lambda/\delta\lambda \sim 5000$ and a heterodyne system with $\lambda/\delta\lambda \sim 10^6$. Although *FIRI* is an ambitious mission, we note that FIR interferometry is appreciably less demanding than at shorter wavelengths.

We envisage *FIRI* as an observatory-class mission because of the vast range of science to be undertaken. We propose that detailed assessment studies in critical areas be undertaken in Europe by ESA, space industry and science institutes, alongside our NASA and CSA partners, with the objective of launching an L-class mission within the Cosmic Vision time frame.

# 1. Introduction

The untapped potential of FIR astronomy is most clearly illustrated by considering the three main components that dominate the electromagnetic energy content of the Universe (Figure 2). The dominant component is the microwave background produced by the primordial Universe at recombination ($z \sim 1089$). The second most important is the FIR background, produced by galaxies in the young Universe (Dole et al. 2006). The third is the optical background dominated by evolved stars/galaxies and AGN. The first and third of these components have now been mapped in detail over the entire sky, while virtually no sky has been imaged in the FIR to any reasonable depth.

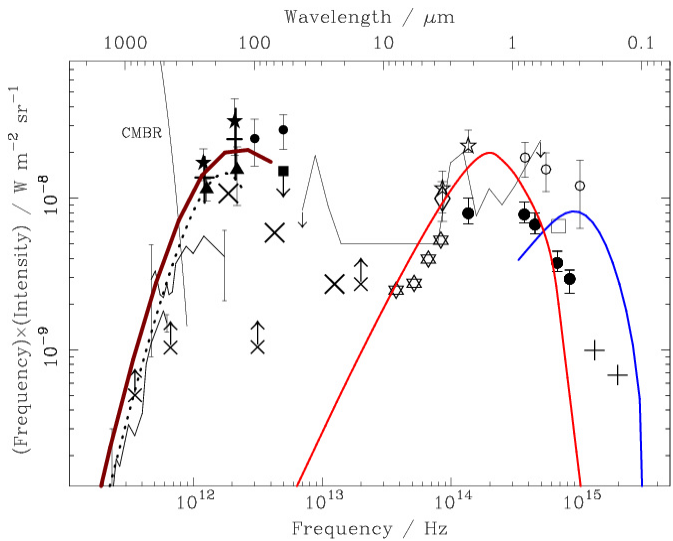

**Figure 2:** Observed UV-to-radio intensity of the cosmic background radiation (Blain et al. 2002). A major goal of observational cosmology is to resolve the entire FIR background. This will probe the earliest formative episodes of stars and galaxies and reveal the effects of feedback on the intergalactic medium (IGM). It will allow us to piece together the transformation of those galaxies through later dust-enshrouded episodes, as well as their links with other galaxy populations selected at X-ray-through-radio wavelengths.

The FIR wavelength region is thus the least explored part of the electro-magnetic spectrum yet it provides uniquely powerful tools to study material associated with the earliest evolutionary stages of galaxies, stars and planets. This waveband allows us to *directly* probe objects during their formation. We gain access to unique science – the peak of the cosmic background radiation and emission from cold, proto-stellar cores.

*1.1 FIRI and ESA's Cosmic Vision 2015-2025*

*FIRI* will have a major impact on two of the four ESA Cosmic Vision themes, both fundamental to our understanding of the formation and assembly of the celestial objects we see today – from stars, planets and life, to galaxies and the Universe.

Theme 1: What are the conditions for planet formation and the emergence of life? *FIRI* is *the* key ESA mission for answering questions about the evolution of gas and dust into stars and planets. Moreover, it will also play a critical role in the search for Earth-like planets that may harbour life, for example via its ability to image the dust structures in these planetary systems. Specific questions posed in the Cosmic Visions program that *FIRI* will be able to answer are:

- What are the conditions for stars to form, and where do they form?
- How do they evolve as a function of their interstellar environment?
- What are the conditions for planets to form around stars?

This third sub-theme is a particular strength of *FIRI*, with its extremely high spatial resolution tracing changes in dust properties and conditions (and water!) across the proto-planetary disks.

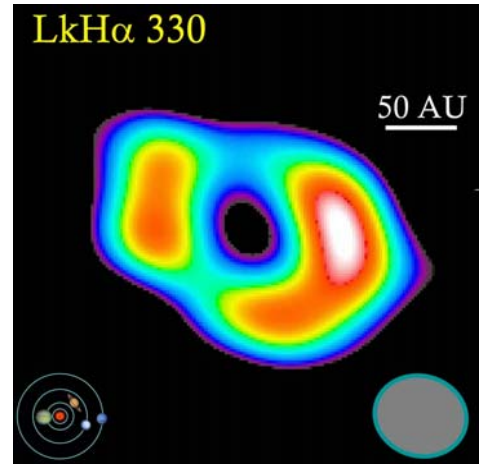

**Figure 3:** Sub-millimetre Array image of the dust continuum emission from the transitional disk around LkHα 330, a G3 pre-main sequence star in Perseus. The putative outer gap radius of ~50 AU, derived from a spectral energy distribution (SED) analysis, is confirmed. Gas is detected well inside the transitional radius, raising the possibility that the gap is caused by a planetary companion.

Theme 4: How did the Universe originate and what is it made of? As noted already, most energy from galaxies in the early Universe is observed at long wavelengths, in the FIR regime. Specific questions set by Cosmic Visions where *FIRI* will be crucial include:

- How were the first luminous objects in the Universe ignited? How did the very first stars form and evolve? How did the history of stars and supernovae give rise to current chemical element abundances?
- What is the history of SMBHs? How do they interact with their host galaxy?
- What is the nature of the FIR background, and of early, deeply embedded star formation?

*1.2 The recent history of FIR astronomy*

*IRAS* is often credited for opening our window on the dusty Universe (Beichman 1987; Soifer et al. 1987), inspiring the line "God lives at 100 μm" from Allan Sandage. Over the course of 300 days in 1983 it surveyed 96% of the sky at 12, 25, 60 and 100 μm, discovering not only Vega-type debris disks (Aumann et al. 1984) and thousands of hot, dense, star-forming cores in our Galaxy, but also a new class of dusty, gas-rich starbursts (ULIRGs – Soifer et al. 1984; Sanders &

Mirabel 1996), which is now known to be the tail end of a more numerous population of luminous, dusty starbursts at higher redshift: submm-selected galaxies (SMGs – Smail et al. 1997; Hughes et al. 1998).

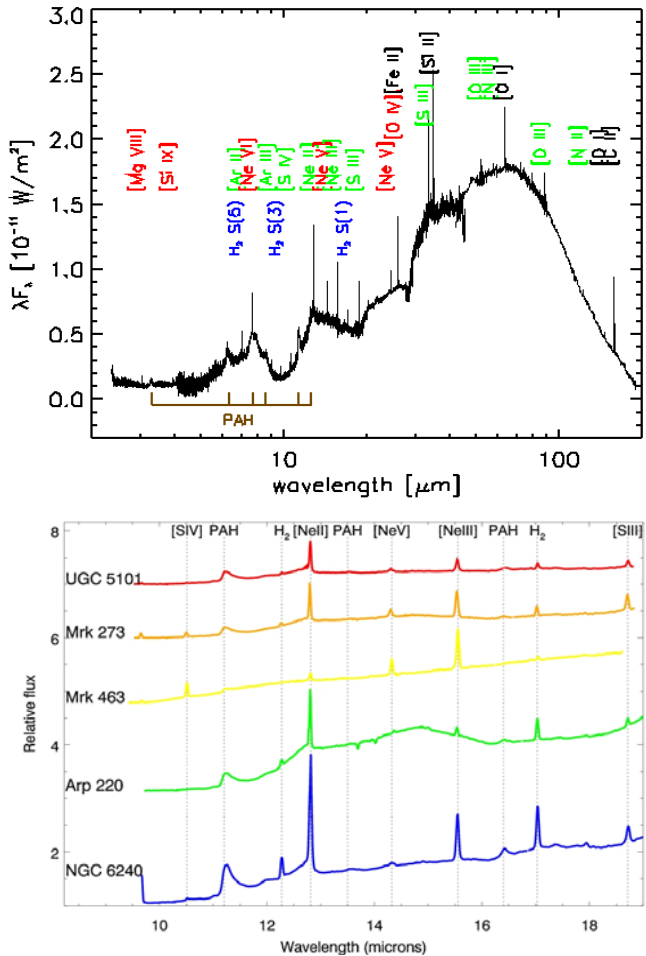

**Figure 4:** *Upper:* 2- to 200-μm *ISO* spectroscopy of Circinus (Sturm et al. 2000). FIR observations can answer fundamental questions regarding the formation and early evolution of galaxies, the source of the extragalactic background at low and high energies, and the origin of nuclear activity in galaxies. Quantitative FIR spectroscopy can be decisive in determining the relative roles of star formation and active nuclei in providing a galaxy's bolometric luminosity. *Lower: Spitzer* MIR spectra of five local ULIRGs (Farrah et al 2007). A variety of key diagnostic features are clearly visible, including PAHs, which signify on-going star formation, the [Ne V] line, which indicates the presence of a buried AGN, and [Ne II] and [S III] which can be used to infer the density and excitation of the narrow-line region gas. *FIRI* will easily detect features such as these, highly redshifted, in extremely distant galaxies.

ESA's cryogenic *Infrared Space Observatory (ISO)* was an unprecedented success, with almost 1000 high-impact refereed papers describing major scientific insights, particularly the discovery of crystalline features in objects ranging from comets to proto-stellar disks, the first detection of FIR water lines, the confirmation of evolution in the LIRG (~$10^{11}$ $L_\odot$) population from $z = 0$ to 1 via deep 15-μm imaging, the discovery of cold (~20 K) dust components in normal, inactive spirals at 175 to 200 μm, and the development of diagnostic ratios via spectroscopy of cooling lines and PAHs (Figure 4; e.g., Genzel & Cesarsky 2000).

Despite pioneering breakthroughs in the FIR regime, *IRAS, ISO* and *Spitzer* have provided only a glimpse of what is to come. Their angular resolution, at best 60 arcsec at 200 μm, makes source confusion a major issue at only moderate depths at the wavelength where the energy density from galaxies is greatest (Figure 5).

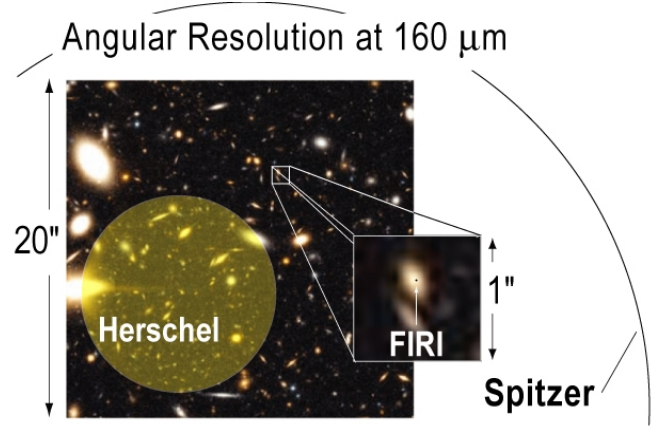

**Figure 5:** A simulated *JWST* deep field, illustrating *FIRI*'s ability to distinguish the emissions of individual galaxies. For comparison, the *Spitzer* resolution at 160 μm is coarser than the entire 20-arcsec field shown, and the 3.5-m diameter *Herschel* telescope will see *many* objects per beam at this wavelength. *FIRI*'s synthesized beam is barely visible (as a black dot) even in the blown-up arcsec² image to the right.

ESA's *Herschel,* scheduled for launch in 2008, is expected to yield a major breakthrough for FIR astronomy. Its resolution at 200 μm, 5× better than *ISO*, will allow imaging photometric and spectroscopic observations of the interstellar medium (ISM) in our Galaxy and its neighbours, and extend the wavelength coverage into the submm waveband. Source confusion, however, will remain an issue. In even relatively short integrations it will be unresolved sources, not photon noise on the detectors, which will limit the depth of an observation. This is true for both extragalactic and Galactic observations, with distant galaxies dominating the noise even in moderately luminous regions of the plane of the Galaxy, an effect already noted by researchers searching for debris disks in *Spitzer* images. This will remain the case for *SPICA*, although *SPICA*'s raw sensitivity will be within 1 to 2 orders of magnitude of *FIRI*, depending on detector developments.

*1.3* The way ahead for FIR astronomy

A large, single aperture in space would provide unprecedented increases in sensitivity and mapping speed over any existing or planned facility, particularly if actively cooled. However, it is clear that to make significant advances on *Herschel* and *SPICA* in all areas of astronomy, from nearby planetary systems to the highest redshift galaxies, even the largest conceivable dish would be inadequate. These areas call for exquisite angular resolution, around 0.02 arcsec at 100 μm, as well as sufficient sensitivity to allow photon-starved spectroscopy across an arcmin² field of view in the 25- to 385-μm band. ***These science requirements lead us inexorably to an interferometer.***

## 2. The FIRI Science Programme

A space-based FIR interferometer will enable the following unique science:

Planets and life

The FIR is a critically important wavelength range for studying the origin and evolution of planetary systems. Most solar systems, including our own, are pervaded by dust, which is very bright at FIR wavelengths. By studying the structure and dynamics of this dust, with many times the effective sensitivity of ALMA, we gain information on how such systems were formed. More importantly, we can infer the presence and orbital motions of planets, which influence the distribution of the dust. Complementary to this, the FIR is the natural place to observe the emission of organic molecules that could literally be the building blocks of life in the Universe. By better understanding the contents and chemistry of the interstellar medium (ISM) throughout the wide range of different environments found in the Milky Way and nearby galaxies, it will be possible to get a better idea about the potential for life in these planetary systems.

Stars and local galaxies

In relatively quiescent local galaxies, including the Milky Way, the bulk of the ISM is relatively "cool", with characteristic temperatures of order a few tens to a few hundreds of Kelvin. Therefore, the chemistry and dynamics of the ISM in these galaxies is often accessible only through FIR observations, as "cool" material radiates most strongly in the FIR. In particular, emission lines of CO offer a powerful tool for investigating dynamic processes in local (and distant) galaxies, and when combined with emission lines from atoms and ions of O, N and C, can be used to build up a picture of the chemical composition of the ISM. Other useful probes of the "cool" ISM include emission and absorption features from solid grains, most notably polycyclic aromatic hydrocarbons (PAHs), and large amorphous silicates.

Conversely, the most luminous sources of radiation in galaxies are either very hot, very young stars, or the accretion disk around the central SMBH. Both hot young stars and accretion disks emit nearly all their energy at UV, X-ray and even γ-ray wavelengths. However, these sources are in most cases embedded in large amounts of gas and dust; for example, the earliest stages of star formation invariably occur deep inside clouds of interstellar gas and dust – their raw material, and AGN require the accretion of large amounts of gaseous fuel onto the central SMBH. This surrounding gas and dust absorbs most or all of the UV and soft X-ray radiation which is directly emitted by the sources, and re-radiates the bulk of it in the FIR; the surrounding clouds are transparent at these wavelengths and the grains can achieve an energy balance between radiation absorbed at short wavelengths and emitted at long wavelengths. As a result, most of the energy originally emitted by the high-temperature primary sources is converted into FIR and submm radiation. Only by observing at these wavelengths can we measure fundamentally important parameters such as total energy budgets ($L_{bol}$), accretion disk geometries and star-formation rates.

Indeed, there is already strong evidence that the growth of the SMBHs found in the centres of nearby galaxies probably takes place at the same time as the formation of the bulk of stars in the central bulge. As a result, this growth likely adds significantly to the energy output of a galaxy in the FIR, but not at any other waveband. FIR observations thus provide the only way to detect this SMBH growth; even hard X-ray observations cannot penetrate the Compton-thick shrouds of gas in a forming galaxy. The synergy between the next X-ray and FIR facilities is obvious.

Cosmology

There is now convincing evidence that the bulk of star and galaxy formation occurred in the first half of the history of the Universe. The FIR wavelength range contains by far the greatest part of the energy output from these galaxies. Thus, FIR wavelengths can provide insight into transformational processes taking place in the Universe from 100 Myr to the present day, resolving the FIR/submm background produced by galaxies in the young Universe (Chapman et al. 2005).

Perhaps the biggest impact of FIR observations will come from studies of the first luminous objects in the Universe, only ~100 Myr after the Big Bang. At these epochs, before the Universe was fully re-ionised, observations at optical/NIR wavelengths are impossible due to the Lyman opacity of hydrogen. FIR data, on the other hand, can provide fundamental new insights into the transformational processes taking place in the ~100-Myr-old Universe. We expect the first stars in the Universe to be extremely massive, metal-poor and very short lived. The earliest formation stages of these stars will be signposted by the FIR-bright rotational-vibrational transitions of molecular hydrogen as it cools and collapses under its own gravity, while the supernova (SNe) explosions associated with the end of the lives of these massive stars will seed the IGM with metals and dust, which will also emit strongly at FIR wavelengths.

The bulk of galaxy formation is thought to occur after re-ionization, between 500 Myr and 6 Gyr after the Big Bang. Nearly all of the stellar and black hole mass build-up in these galaxies occurs while shrouded in dust, making FIR observations essential to our understanding of how these galaxies are forming. Though high-redshift galaxy surveys in the optical have detected many distant galaxies, they are virtually all systems that have already built up a significant fraction of their stellar mass. At FIR wavelengths we can probe both the direct formation phase, and the degree of on-going star formation embedded deep in giant molecular clouds, to provide a view of galaxy formation that is orthogonal to that taken in optical/NIR surveys.

FIR telescopes probe the peak of the SED of galaxies so *FIRI* will reveal, being immune to the effects of extinction, the links between morphology, environment, gas content, metallicity and the decline in specific star-formation rate that has occurred since $z \sim 1$, commonly referred to as "downsizing". By tracking the total luminosities of galaxies, all the way to the present epoch, the true form of evolution of the luminosity function of galaxies will be revealed and we can be sure for the first time about the systematic change in opacity, luminosity and extent of star-forming regions in galaxies as a function of their environment, mass and luminosity. Without direct FIR imaging, these quantities can be inferred only by proxy.

In what follows we discuss the likely scientific impact of the FIR interferometric mission, *FIRI*, comprising three cold, 3.5-m apertures, orbiting a beam-combining module with a maximum separation of ~1 km, free-flying or tethered, operating between 25 and 385 μm, using the interferometric direct-detection technique to ensure μJy sensitivity and 0.02-arcsec resolution at 100 μm. It will revolutionise our knowledge of the formation of galaxies, stars and planetary systems and the development of life-sustaining environments. We will be able to probe the universality of the initial mass function across a range of galaxy environments and map out star- and planet-forming disks in stellar nurseries through resolved spectral lines. *FIRI* will break the cosmic background radiation into its constituent parts – many thousands of faint, dusty high-redshift galaxies, resolving them individually to yield otherwise hopelessly obscured information about their formation and evolution. It will root out Compton-thick AGN and differentiate between gas heated by active nuclei and stars, thus disentangling the formation histories of SMBHs and stars. In summary, *FIRI* is capable of answering many of the most important questions posed by Cosmic Vision.

All the science topics in this proposal can be addressed with *FIRI*. One should, however, not underestimate the power of serendipity. Many of the mission highlights already discussed were not predicted or expected. An orders-of-magnitude increase in angular resolution coupled with extreme sensitivity will undoubtedly open fields in astrophysics that are currently unknown.

2.1 Introduction to planet and star formation

One of the most fundamental goals in astrophysics is to understand the formation of stars and planets. This topic has been the target of numerous studies, especially over the last three decades. The advent of IR and submm instrumentation boosted this research because astrophysicists finally managed to look through the shrouds of dust in which the forming stars are hidden. However, inherent to long wavelengths is a lack of angular resolution. Only recently 8-m class IR-optimized telescopes have become available, while interferometry has only been a recent development in the IR and submm. The interesting scales, however, are very small. The nearest regions of star formation are ~100 pc away. To determine properties that lead to Earth-like planets, we have to look at scales of 1 AU. As a rule of thumb, the interesting angular resolutions are thus at ~0.01 arcsec.

The complex, non-linear physics controlling star and planet formation leads to multiple outcomes depending on variations in the controlling parameters (e.g. environment) but also on random fluctuations. Therefore to build a comprehensive, complete theory requires a complete description of the critical stages: proto-stellar collapse, disk formation and disk clearing. Key points are missing in disk theory:

- What is the origin of turbulence in disks? What roles do magnetic fields play?
- What is their thermal structure? How much heating is contributed by viscous (turbulent) processes?
- How do grains evolve in disks? What impact does this have on disk properties and thermal balance?
- What role does water ice play in grain evolution? What effect does water vapour have on thermal balance?

These questions cannot be solved independently as the gas chemistry determines the fractional ionisation, which impacts the gas coupling to the magnetic field. Viscous heating is an important heating source in disks, which complements the stellar (UV and X-rays) and chemical ($H_2$ formation) heating and can even dominate close to the star. Viscous heating is provided by gas friction and is therefore linked to the level of turbulence in the disk. Furthermore, recent work shows that the disk structure regulates its ability to form planets, and their migration rate. Type-I migration is caused by non-zero torque resulting from planet-disk interaction. This torque is generally negative (causing inward migration) but its level depends on the detailed disk properties, with temperature structure playing an important role.

As for interstellar gas, the physical and thermal structure of proto-planetary disks depends on their chemical composition, which determines the main cooling processes. In their outer regions, it is valid to use PDR models developed for the ISM and adapted to the specific conditions of disks. Calculations have been performed by various groups showing that familiar fine-structure lines, [C II] and [O I], are expected to show up in the disk atmosphere at $>10^{-8}$ W m$^{-2}$ sr$^{-1}$, near 100 AU. Because of the steep temperature gradient [O I] decreases faster than [C II] with distance from the illuminating star in the region inside 100 AU, [O I] cooling being especially important near 10 AU. Other important coolants for the inner regions are fine-structure lines of S, Si and Fe, molecular lines (CO ro-vibrational and rotational lines; $H_2O$ and OH lines), Ly α and the 630-nm oxygen lines, $H_2$ ro-vibrational lines, gas-grain collisions, with minor contributions from other species. In these calculations, the disk atmosphere can become very hot (several thousand K), so predictions from various models may differ substantially depending on small variations in physical processes. An indirect confirmation of the warm region has been provided by the observation of ro-vibrational CO emission lines in a few proto-stellar sources.

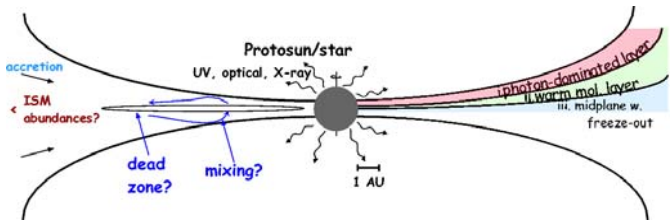

**Figure 6:** Cartoon of a proto-planetary disk.

For all systems, *FIRI* will be able to simultaneously probe the cooling processes and the gas dynamics which will be inaccessible to ALMA and *JWST* (the most interesting water and OH lines are outside the wavelength range studied by MIRI, and its spectral resolution does not yield gas dynamics). In 2015, *JWST* and ALMA will have been operational for several years, sensitive to the coldest areas of proto-planetary disks, i.e. the disk mid-planes. While this provides important information about processes happening in molecule-starved gas, it will not tell us what happens to one of the most important building blocks of life – water: telluric water will block these lines efficiently. Similarly, *JWST* will only be able to probe the highest rotational transitions of water, giving access to the more exotic phenomena (like shocks) in these disks. Only the FIR region gives access to the most important water lines.

Evaporation caused by external radiation is another process limiting planet formation. Proplyds in the Orion nebula are among the best examples of proto-stellar disks exposed to intense FUV radiation. The perturbed morphology of these systems is a clear sign of the role of environment on disk morphology. In these systems, a photo-dissociated region (PDR) develops at the disk surface. Strong fine-structure line emission is produced in this PDR, which allows us to make a complete inventory of the disk structure and chemistry. The dust that causes the silhouette in the well-known *HST* images can be well characterised using *FIRI*'s spatial resolution. Determining the dust temperature via *FIRI* photometry will be a major step forward.

Some other lines also deserve closer inspection, e.g. the 63-μm [OI] line, high *J*-lines of CO and the ground-state rotational lines of HD and OH. These, together with water, give an excellent census of oxygen and thus provide a range of probes for physical processes and chemical conditions which cannot be obtained in any other way.

While gaseous water is an important ingredient in the planet-formation process, the dust is also playing a major role. After all, most planets in our own Solar System are (partly) rocky. *JWST*'s MIRI will give access to detailed dust studies, but it has no access to the wavelength region where dust temperature can be determined accurately. The 30- to 150-μm region offers this possibility in the form of a temperature-sensitive water-ice band. It is not only proto-planetary disks that can be followed this way; even objects in later stages of disk evolution, like debris disks, can be probed efficiently, given sufficient sensitivity. In these debris disks the dust in the systems is created by collisions of rocky bodies or by cometary dust-tails. *FIRI* offers the chance to detect the equivalent of Comet Hale-Bopp in an extra-solar system.

*FIRI* studies of dust and gas, supplemented with *JWST* data and ALMA/EVLA interferometry, will provide the detail needed for a colossal leap forward in our understanding of planet- and star-formation processes.

## 2.2 Planet formation and the role of water

The conditions in the disks that surround newly formed stars – the birth sites of future planetary systems – run through an even more extreme range of temperatures and densities than can be found in interstellar clouds. From the dense and cold mid-planes, temperatures rise and densities drop with increasing height in these flared disks. With decreasing distance to the central star, temperature also increases from ~10–20 K at a few hundred AU to hundreds of K within 1 AU. The UV and X-ray radiation from the active pre-main sequence star, or from neighbouring young stars, strongly affects the chemistry of the disk gas, evaporating or desorbing molecules from the ice mantles of dust grains into the gas phase, but also dissociating many species.

These complex environments in protoplanetary disks suggest that water is likely to be present with a large range of abundances (four orders of magnitude or more) within each disk, as encountered throughout the ISM of the Galaxy. Water lines can therefore probe the gas distribution within the disks. Furthermore, the presence of water is highly relevant for the formation of planetary systems. The leading theory for the formation of gas-giant planets involves the presence of water ice on dust particles: icy surfaces stick better to each other, helping planetesimals to grow and accumulate, until proto-planetary cores have sufficient gravity to capture a large $H_2$ envelope. Closer to the star – inside this "snow line" – where temperatures are higher and water is expected to be in the gas phase, only smaller, rocky planets like the Earth should form. However, around 1 AU the presence of water is vital for the possible formation of oceans on terrestrial planets. Water may be accreted locally if trapped inside porous dust grains, or it may require to be "delivered" from impacts of acqueous asteroids or icy cometary bodies. Observing the water content of disks can show from where exo-Earths acquire their oceans, and how important random events may be: the different $H_2O$ content and D/H ratios on Earth and Mars suggest just a few impact bodies can determine the nature of the future oceans.

Resolving the spectral lines of water with high velocity and angular resolution is important. In the "baseline" chemical model, one expects water to be largely frozen out throughout most of the disk volume. However, theories of planet formation suggest that when planetary cores form, their gravitational pull can open gaps in the disk or excite spiral waves. These features may have a profound effect on the local water gas-phase abundance, and leave an imprint on the shape of spectral lines that would pinpoint the location of planet formation. A FIR interferometer will be uniquely sensitive to the warm gas-phase water in regions of disks akin to the inner Solar System. In contrast, longer-wavelength facilities such as ALMA are sensitive mostly to cold gas, as in outer cometary zones, and will struggle to detect the relatively high-frequency lines of water and its isotopes against strong atmospheric absorption.

All the above-mentioned aspects make a strong case for high spectral and angular resolution observations of water lines. A simple predictive model can be constructed starting with disks consistent with the Taurus star-forming region (d'Alessio et al. 1998). The results show that the strongest lines are among the lowest transitions (e.g., $1_{10}$–$1_{01}$, $2_{12}$–$1_{01}$, and $1_{11}$–$0_{00}$). Typically, these lines are 0.5 to 1.0 K in a ~0.02-arcsec beam. The emitting region is about 0.1-arcsec in size, so smaller beams start to resolve the emission: the Habitable Zone of the Earth's orbit fills a 0.02-arcsec beam 100 pc away. Several nearby young star clusters are closer, so direct imaging of the formation zone of exo-Earths is possible. The emission is dominated by the >200 K gas where the water abundance is high, and very well matched to Earth-like temperatures of ~280 K.

It should be stressed that this "baseline" model does not include any effects of gaps, spiral arms, shocks or UV radiation that may penetrate the disk. For example, results by Ceccarelli et al. (2005) and Dominik et al. (2005) suggest that increased photo-desorption may increase the water gas-phase abundance for even weak radiation fields. Any regions of enhanced density or temperature due to shocks or gaps on order 1-AU scales, with increased water gas-phase abundance, would significantly increase the line strength in 0.02–0.1-arcsec beams.

2.3 The road to solid bodies such as asteroids and rocky planets

At the time of writing, over 200 planetary objects orbiting Solar-type stars have been confirmed, down to as little as 5 $M_\oplus$. These discoveries show that planet formation is common in the Universe. To derive the mechanism leading to the formation of a planetary system one needs to observe the circumstellar disks from which the planets are born. Current understanding has identified the circumstellar disks surrounding young (1 to 10 Myr) low-mass (0.1 to 1 $M_\odot$) T Tauri stars (TTS) and intermediate-mass (~2 $M_\odot$) Herbig Ae/Be stars (HAEBE) as the sites of ongoing planet formation ("planet-forming disks").

By studying their solid state and gaseous components, the physical conditions and processes leading to the formation of planets inside these disks can be determined. A substantial fraction of these circumstellar disks have masses and sizes comparable to the expected values for the primitive Solar nebulae, and so are natural candidates for the birth sites of planets like the Sun's (Strom et al. 1989; Beckwith et al. 1990; McCaughrean & O'Dell 1996; Dutrey et al. 1996). The sub-μm sized dust grains present at the formation time of these disks can coagulate and form larger objects and eventually Earth-like planets (e.g. Weidenschilling 1997). The formation of giant gas planets can either follow the formation of a massive rocky core and a subsequent fast gas accretion (Pollack et al. 1996) or alternatively through gravitational instabilities and subsequent fragmentation of a massive disk (e.g. Boss 2001).

Though current instrumentation has provided us with the first insights into the workings of these planet-forming disks, it is still limited by angular resolution and/or sensitivity. Of the instrumentation currently under construction, ALMA will mark a major advance in our understanding of planet-forming disks. With its sensitivity and resolution, it will allow us to spatially resolve the disk structure in great detail. However, dust associated with the formation zones of the Earth and Jupiter, at ~100 to 300 K, will emit far more strongly in the FIR. Measuring the radial dust distribution is vital to understanding the architecture of future planetary systems – for example, Kuchner (2004) infers surface density profiles of $r^{-2}$ for multiple exo-planet systems, crowding the planets much closer to the star than the $r^{-1.5}$ of the primitive Solar nebula. Measuring the dust distribution close to the star is very important for understanding if the Solar System architecture is expected to be common or rare.

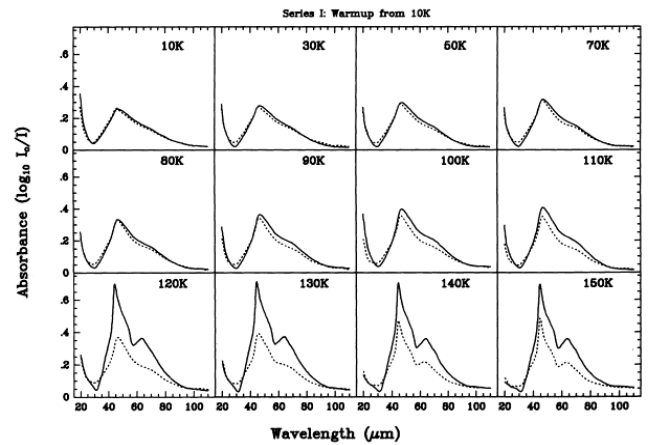

**Figure 7:** FIR absorbance spectra of thin $H_2O$-ice film (*solid*) and thick $H_2O$-ice film (*dotted*), measured at the temperatures indicated beside each spectrum (Smith et al. 1994) showing the shape and frequency changes of the water ice band.

Submm observations will not provide us with details on the chemical composition of the solid-state components. At shorter wavelengths, *JWST* and especially its MIRI instrument will provide breakthroughs, especially on crystalline compounds; however, it has no access to the water ice bands at 44 and 62 μm (Figure 7) and it also lacks angular resolution at its longest wavelengths. To study the dust chemistry in the formation regions of asteroid and Kuiper-belt structures and Jupiter-like planets, the FIR waveband provides most information. As these wavelengths are not accessible from the ground, *FIRI* is required.

The ultimate goal is to identify the physical processes determining the disk structures and chemistry, leading to a complete understanding of the workings of planet-forming disks. The initial conditions and timescales of planet formation can be observationally determined, with the possibility of witnessing the formation of a planetary system directly through the influences it exerts on the disk structure and chemistry. The FIR waveband can uniquely identify the dust composition of all major crystalline silicate components formed in the disks, and so the base material for planetary compositions. Further, it is the only spectral window for studying species like water ice, carbonates and hydrated silicates. Water ice chemistry traces the formation region of Jupiter-like planets and hydrated silicate species the processing of materials in protoplanetary bodies. With such observations, a direct comparison can be made with with observations of comets, asteroids and meteorites – all preserving a record of the early evolution of the Solar System – and our understanding of the formation of our planetary system can be enhanced greatly.

2.4 Determining the origin of the Initial Mass Function in the Milky Way and beyond

The "initial mass function" (IMF) – the distribution of stellar masses at birth – is one of the fundamentally important parameters in astrophysics. Its origin is thought to lie in the detailed physics of star-forming regions and their turbulent disintegration into proto-stellar cores (e.g. Bonnell, Larson & Zinnecker 2007). It has been argued both that the IMF should be universal (e.g. Baldry & Glazebrook 2003) and that the IMF is different in specific classes of objects such as starbursts (e.g. Rigby & Rieke 2004) and primeval galaxies. Recent observations of submm galaxies (Smail et al. 1997; Hughes et al. 1998), for example, can only be explained in some models by positing a low-mass cut-off in their IMF (Baugh et al. 2005).

As pre-stellar cores and young (Class 0) proto-stars emit the bulk of their energy longward of 100 μm (André et al. 2000; Figure 8), FIR continuum mapping is a unique tool to measure the IMF. Recent ground-based continuum surveys of nearby compact cluster-forming clouds (ρ Oph, Orion B) at 850 and 1200 μm have uncovered "complete" (but small) samples of pre-stellar condensations (i.e. starless cores) whose associated mass distributions resemble the IMF (Motte et al. 1998), albeit sampling their SED at a point far removed from the peak. These findings suggest that the IMF of solar-type stars is largely determined by pre-collapse cloud fragmentation, prior to the proto-stellar accretion phase. They are, however, seriously limited by small-number statistics at both the low- and high-mass ends of the spectrum due to both sensitivity and angular resolution limitations of current telescopes.

In our Galaxy, *Herschel* will determine the pre-stellar core mass function from the proto-brown dwarf (M ~ 0.01 M$_\odot$) to the intermediate-mass (M ≤ 8 M$_\odot$) regime, with good statistics, via an extensive survey of the nearest (< 0.5 kpc) molecular cloud complexes of our own Galaxy, mostly belonging to the Gould Belt (André & Saraceno 2005). This study will, however, be limited to nearby regions because its 15-arcsec resolution (at ~200 μm) is only able to probe individual (~0.05 pc) star-forming cores out to 0.5 kpc. Unfortunately, this volume of Galactic space will yield an insufficient number of massive pre-stellar cores and it will not be possible to investigate the core mass function in the regime of OB stars (M$_{star}$ > 8 M$_\odot$).

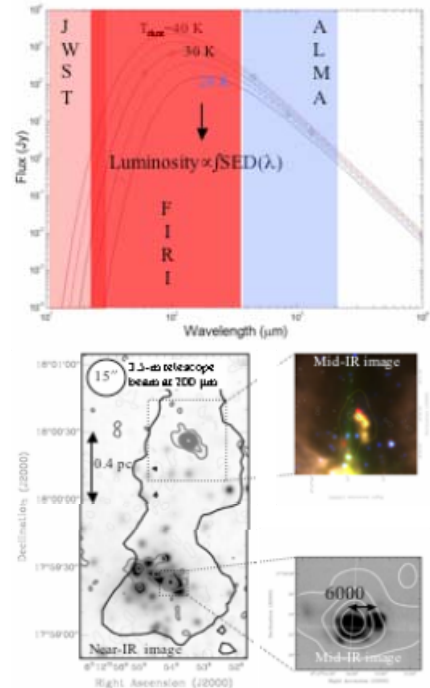

**Figure 8:** *Above:* Likely SEDs for the early phases of high-mass star formation alongside the observed SED of a massive proto-stellar core at 5 kpc. Data at ~60–400 μm will constrain the SED peak, especially for cold sources (10–50 K), and hence tie down temperature and luminosity. *Below:* Typical high-mass star-forming clusters embedded in a larger star-forming complex (Minier et al. 2005). Black contour: cold dust submm continuum emission, which defines two clumps; greyscale image: NIR (2.2 μm, ~1-arcsec resolution); small, grey contours: free-free continuum emission at 15 GHz (~1-arcsec resolution). The southern and northern clumps harbour a cluster of embedded young massive stars emitting in the NIR, and in radio continuum for the most ionising ones. Young massive (proto-)stars are not detected in the northern clump via NIR emission, but show up at radio and MIR wavelengths. For the earliest phases of the clustered star-formation process, during which extinction is too high in the NIR/MIR, sub-arcsec FIR observations are necessary to derive the luminosity and mass of the most embedded proto-stars down to binary system scales (<100 AU).

Despite their rarity, massive stars (M$_{star}$ > 8 M$_\odot$) dictate the evolution of galaxies. They impact their environment through their strong ionizing radiation, powerful stellar winds and violent deaths. In addition, massive stars are the primary source of heavy elements. Heavy elements from C to Fe are synthesized during their short lifetime while heavier elements are produced by SNe by neutron capture. Metal-poor environments show an excess of extremely massive stars whereas metal-rich environments produce lower-mass stars. The overall process of isolated low-mass star formation is now well understood (e.g., Shu et al. 1987). However, most stars, and especially high-mass stars, are presumed to form in clusters (e.g. Adams & Myers 2001). In contrast to low-mass star formation, the basic processes of high-mass star formation are still largely unknown and poorly constrained by observations. It has been suggested by Bonnell, Larson & Zinnecker (2007) that massive stars form through proto-stellar mergers which would be common at the centre of forming clusters. Alternatively, high-mass stars may form directly through collapse if the typical accretion rates are several orders of magnitude larger than expected, and large enough to overcome radiation pressure (e.g. McKee & Tan 2002). In fact, when a forming star reaches 10 $M_\odot$ the accreting flow is believed to be completely stopped by the radiation pressure of the ionising star. In addition, it is not known whether clusters of hundreds of low- and a few high-mass stars form from assemblies of decoupled molecular fragments or from competitive accretion inside a large, massive infalling cloud (Padoan et al. 2001; Bonnell et al. 1997). Whatever the scenario for high-mass star formation, specific predictions can be tested observationally with *FIRI*.

### Star formation in the Local Group: how universal is the IMF?

It is well known that the star-formation rate per cubic Mpc (co-moving) decreases by more than an order of magnitude between $z = 1$ and now (roughly half the Universe's lifetime ago), e.g. Madau et al. (1996); Heavens et al. (2004). There was certainly not 10–30× more molecular gas present back then, so this suggests that star-formation rate per unit $H_2$ (the star-formation efficiency, SFE) was significantly higher than now. The star-formation rate today is dominated by large spiral galaxies, but this was much less true in the past when spirals were smaller (at least their stellar part), bluer, more gas-rich, and less chemically enriched.

These differences show how important it is to sample star formation in a wide range of environments and Local Group galaxies provide an excellent test-bed. The Magellanic Clouds, with metallicities ~1/3 and ~1/5 Solar, and masses ~1/10 and ~1/50 of the Milky Way, are only 55 and 70 kpc away. They also have a higher gas-to-stellar mass ratio than the Milky Way or Andromeda. Recent studies indicate that the Local Group spiral, M33, has an unusually high SFE, although the cause is not clear. The Local Group Irregular, IC 10, apparently shows the same characteristic, although there is a greater uncertainty associated with the $N(H_2)/I_{CO}$ ratio (Leroy et al. 2006).

*FIRI*'s angular resolution will allow us to study proto-stellar condensations in the Magellanic Clouds, in a manner analogous to studies of our own Galaxy with *Herschel*, but probing the IMF over a larger range of mass. Studying star formation in the Galaxy provides access to how stars form in a roughly Solar-metallicity environment situated in the gravitational potential of a large spiral with the associated differential rotation, epicyclic motions, shear, etc. Furthermore, the stellar surface density dominates over the gas out to quite large radii, which could have an effect on the structure of molecular clouds, the efficiency of star formation, and thus potentially the IMF.

At 55 kpc, *FIRI* probes sizes down to 1000 AU, small enough to study sub-Solar mass condensations. Assuming a gas-to-dust mass ratio a factor of a few higher than the Milky Way, due to the lower metallicity, we expect 100-μm flux densities within a 0.02-arcsec beam of 100 μJy for dust at 20 K (or 10 μJy at 15 K). Fluxes are higher at 200 μm, although twice the telescope separation is required for the same angular resolution. This is the crucial wavelength regime because most of the energy of proto-stellar condensations is emitted at wavelengths shorter than can be feasibly detected with ALMA (even being *very* optimistic). *FIRI* thus offers the only method of measuring temperatures and luminosities. The flux densities mentioned earlier allow the study of masses as low as 0.5 $M_\odot$ in the LMC and 0.8 $M_\odot$ in the SMC (allowing for the lower metallicity and greater distance). The H II regions in the Magellanic Clouds provide examples of much more active star formation than anywhere in the Milky Way and probing such environments is only possible with *FIRI*.

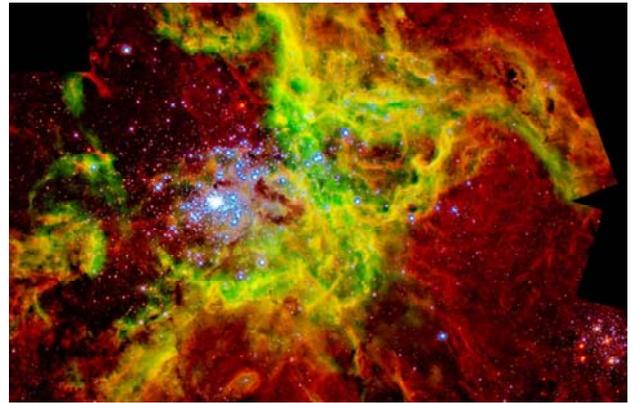

**Figure 9:** *HST* mosaic of 30 Doradus – the most massive star-forming region in the Local Group of galaxies – taken in F814W, F555W, F336W, F673N ([S II]) and F656N (Hα). 30 Dor is located in the LMC, one of our closest galactic neighbours. Red: low-excitation, where gas is neutral or singly-ionised; yellow/green: high-excitation areas, where almost all hydrogen is ionized and other gases (e.g. S, O) are doubly ionised (Walborn, Apellániz & Barbá 2002).

While even *FIRI* will not be able to study the IMF in primeval galaxies, it *will* be capable of studying it in a range of nearby environments, including starbursts. By exploiting its exquisite sensitivity and angular resolution we can improve on studies of the IMF in our own Galaxy and test its universality. To date, the IMF in external galaxies has not been observed directly; instead, it has been inferred by indirect means. *FIRI*, uniquely, allows us to measure the clump mass function directly. *FIRI* will be able to resolve and detect individual pre-stellar clumps in external galaxies. By mapping out star-formation regions in a range of objects the clump mass function can be constructed.

In the Galaxy, the luminosities of proto-stellar condensations vary roughly as $L \sim M^3$. If true for other environments, this would allow masses down to 2 $M_\odot$ in NGC 6822 (490 kpc) or about 3 $M_\odot$ in M33 (840 kpc), for metallicities similar to that of the LMC. Higher angular resolution would make such studies even more powerful. M 33 is a true spiral galaxy with 10% the mass of the Milky Way and a metallicity 2–3× below Solar, likely very typical of intermediate-redshift galaxies. Its star-forming regions are a similar size to those in the Galaxy – is the IMF of the proto-stellar cores similar too? Could preferential formation of high-mass stars explain the apparent high efficiency of star formation (measured by H α, far-UV, MIR emission) in M 33 and the intermediate-redshift population? These are fundamental questions that can only be answered with a FIR interferometer.

2.5 The origins of dust

The puzzling origin of dust in the early Universe has been highlighted by submm observations showing that QSOs at $z > 5$ contain $10^8$ to $10^9$ $M_\odot$ of dust (Bertoldi et al. 2003; Beelan et al. 2006). It is unlikely that this dust could have formed in the winds of low- to intermediate-mass stars (Asymptotic Giant Branch – AGB) since these stars evolve slowly before reaching their dust-producing phase and the time available is < 1 Gyr (Morgan & Edmunds 2003). A possible solution to this problem is that Type-II SNe could also be significant sources of dust. Their rapid evolution means that they could be the dominant dust factories in the early Universe. Theoretical models predict that Type-II and PISN SNe should convert 2 to 30% of their progenitor's mass into dust (Todini & Ferrara 2001; Nozawa et al. 2003; Schneider et al. 2004; see also Sugerman et al. 2006 and Meikle et al. 2007 regarding core-collapse SNe). Dust enrichment from primordial population III SNe is a vital ingredient in the formation and evolution of the first population II stars and galaxies as grains play an important role in cooling the gas and in the formation of $H_2$ (Schneider et al. 2006; Cazauz & Spaans 2004). An observational test of these models is therefore imperative.

While we cannot directly observe population III stars, we can observe present-day SNe to see if they produce the quantities of dust predicted: 0.4 to 1 $M_\odot$ for a 20-$M_\odot$ progenitor. Repeated searches for such dust in young Galactic SNR using *IRAS*, *ISO* and *Spitzer* have been fruitless, with only tiny quantities ($10^{-6}$ to $10^{-3}$ $M_\odot$) found (Arendt et al. 1999; Douvion et al. 2001; Bouchet et al. 2006). One possibility for the failure is that the dust is too cold to emit at MIR wavelengths, while the longer wavelength detectors on *ISO* were confused by foreground dust. Two young remnants were observed with SCUBA to see if this was the case and a large excess of submm emission was found (Dunne et al. 2003; Morgan et al. 2003), spatially correlated with emission at other wavelengths (though the angular resolution of the submm observations was very-much poorer than the optical, radio and X-ray data). These results were interpreted as evidence for significant dust production by SNe but there have been counter-arguments claiming the excess submm emission could be due to a tiny mass of iron needles (Dwek 2004). Also, observations of molecular gas tracers towards Cas A suggest that a large fraction of the submm emission may be due to the foreground ISM (Krause et al. 2004; Wilson et al. 2005). Thus the jury is out on this issue.

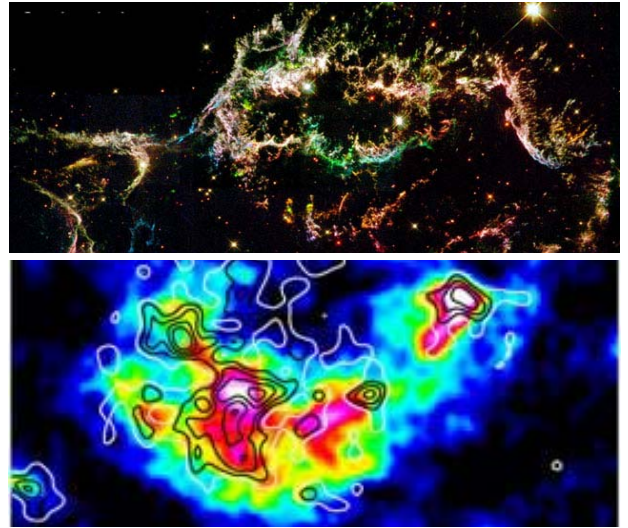

**Figure 10:** Upper: Cas A as observed by *HST* at sub-arcsec resolution. Lower: Cas A at 850 and 450 μm, as observed by SCUBA at 16-arcsec resolution (Dunne et al. 2003).

To determine unambiguously how much dust is produced in SNe requires:

- high spatial resolution, so that we can relate the dust emission to that at other wavelengths which are known to originate within the remnant (radio synchrotron, X-ray and optical). Current submm/FIR instruments, with angular resolutions of 8 to 40 arcsec, make it very difficult to determine the nature of the emission and separate out foreground contamination;

- broad wavelength coverage, to track both the warm and cold dust components. Warm dust emits out to ~100 μm; beyond this is a dip, followed by a rise at > 200 μm. *FIRI* thus captures both peaks and will suffer little or zero contamination from synchrotron. ALMA will extend the SED to longer wavelengths, helping to measure dust properties and map the surrounding cold molecular gas. The velocity structure of the cold gas (via ALMA) will discriminate between material in foreground clouds and material in the SNR; spatial comparison of this with the emission traced by *FIRI* can thus determine the origin of the dust;

- the spectral capabilities of *FIRI* allow us to see how conditions in the gas vary across the remnant and relate this to the differences in dust emission (temperature/grain size/mass etc). MIR spectra will provide important diagnostics for the grain populations and how these change with environment in the remnant (e.g. are certain species destroyed preferentially). At sub-arcsec resolution we can trace material through the reverse and forward shocks, comparing properties before and after. We may also find material re-condensing, post-shock. Water lines can help to follow the shocks accurately, as learned from *SWAS* and *ODIN*. HD has been detected in SNRs (Neufeld et al. 2006) and will probe the formation of $H_2$.

These are questions of paramount importance which only *FIRI* can address.

2.6 Introduction to distant galaxies and AGN

The last decade has seen great advances in our understanding of the development of large scale structure from the small density fluctuations present in the cosmic microwave background. Indeed, n-body simulations may now provide a realistic picture of the development of the underlying dark matter distribution (e.g. Springel et al. 2005). However, the physics of the baryonic component of the Universe (i.e. what we actually observe) is far more complex. In the currently-favoured hierarchical paradigm, massive galaxies are the product of generations of galaxies, starting with the first metal-free dwarfs, that have merged over cosmic time as their dark matter halos fell together. Within the first and all subsequent galaxies, stars have synthesized metals, and as these stars exploded as SNe or shed their outer envelopes they have polluted the ISM and IGM with metals in both gaseous and dusty forms.

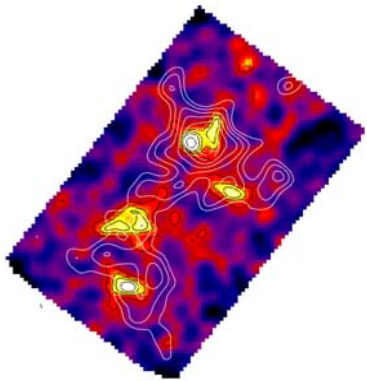

**Figure 11:** SCUBA imaging of the ~1-arcmin$^2$ field around an absorbed $z = 1.8$ QSO revealing a remarkable ~400-kpc-long chain of galaxies, each with an obscured star-formation rate sufficiently high to build a massive spheroid in <1 Gyr (450-μm image; 850-μm contours – Stevens et al. 2004). The genesis of spheroids is central to our understanding of galaxy formation: they are relatively simple systems containing about half the stellar mass of the Universe. A major subset – massive elliptical galaxies – are preferentially found in clusters where they exhibit old coeval stellar populations suggesting that they formed synchronously at early epochs. The over-density of galaxies relative to expectations for a random field implies they reside in a structure associated with the QSO. This star formation is probably associated with galaxy mergers, or encounters within a filament, such as those predicted by hierarchical models. These observations suggest that strong absorption in the X-ray spectra of QSOs at high redshift may result from a veil of gas thrown up by a merger or merger-induced activity, rather than an orientation-dependent obscuring torus. It is possible that these systems are the precursors of elliptical galaxies found today in the core regions of all rich galaxy clusters.

In the hierarchical paradigm, elliptical galaxies result from the merging of disk galaxies over a large fraction of the Hubble time (Figure 11). Within the merging galaxies, gas is ejected, converted into stars, or channeled to the nuclear regions to be accreted by a central SMBH (e.g. Kauffman & Haehnelt 2000; Malbon et al. 2006; Hopkins et al. 2007). The in-spiral and coalescence of the SMBHs of the individual merging galaxies may also be important to the growth of the SMBH. In either case, the growth of the galaxy and that of the black hole is prolonged and coeval.

However, this scenario runs into severe problems when compared with the increasing evidence for "cosmological downsizing" or anti-hierarchic growth of baryonic structures. An alternative view, developed with the specific aim of reproducing the latter phenomenon (Granato et al. 2001, 2004; Ciotti & Ostriker 2004; Silva et al. 2005; Lapi et al. 2006), envisages a very fast collapse of baryons into stars in large galactic halos at high redshift ($z > 2$), promoting the development of a SMBH during a huge star-forming phase lasting 0.5 to 1 Gyr, traced by submm galaxies (Smail et al. 1997; Hughes et al. 1998). In this case, we would expect that the SMBH reaches its final mass toward the end of the star-forming phase, with accretion terminated by strong AGN-related feedback arising when the SMBH is mature enough: the AGN expells the galactic ISM during the last few *e*-folding times of accretion, before the optically bright QSO phase (Granato et al. 2004). FIR observations with high sensitivity and angular resolution are required to test these model predictions, as outlined in what follows.

2.7 The origin of dust through the cosmic ages

Dust plays a crucial role in the evolution of galaxies. On the one hand, dust is known to affect the formation of massive stars. On the other hand, dust is a powerful coolant of molecular clouds, even in low-metallicity environments, and therefore it is expected to facilitate the formation of low-mass stars in primordial galaxies (Schneider et al. 2003, 2006). Dust grains also greatly enhance the formation of molecules, which are also a powerful coolant of the ISM, and therefore catalyse the formation of stars. As a consequence, the amount of dust in galaxies – especially during the early phases of their formation – is a crucial parameter that regulates the rate and mode of galaxy evolution. Moreover, the presence of dust in distant galaxies determines their observational properties and detectability, both through the effect of extinction in the optical-UV and through thermal emission in the IR.

Despite this important role, the mechanism(s) for dust formation through the cosmic ages are still poorly understood, as discussed in Section 2.5. The standard scenario ascribes the bulk of dust formation in local galaxies to the envelopes of evolved stars, AGB stars in particular. Stellar evolutionary timescales constrain the bulk of dust production through the AGB channel to occur at least ~1 Gyr after the onset of star formation. As a consequence, at $z > 5$ (age of the Universe, < 1 Gyr), one expects little or no such dust. However, during the past few years mm/submm observations have detected large masses of dust in QSO host galaxies up to $z = 6.4$ (Bertoldi et al. 2003; Priddey et al. 2003). This dust cannot have been created by AGB stars.

Alternative dust factories are core-collapse SNe: a few hundred days after explosion their ejecta reach a temperature and density appropriate for dust formation (e.g. Todini & Ferrara 2001; Nozawa et al. 2003). Since SNe occur just a few Myr after the onset of star formation, they could provide a fast dust-production mechanism during the early universe. Although the formation of dust in SNe is observationally proven beyond any doubt (Moseley et al. 1989; Spyromilio et al. 1993; Sugerman et al. 2006), the efficiency of dust production is still a hotly debated issue (Dunne et al. 2003; Krause et al. 2004; Wilson & Batrla 2005). In particular, it is unclear whether dust from SNe can account for the quantities observed in high-redshift QSOs (Maiolino et al. 2004, 2006; Dwek et al. 2007; Bianchi & Schneider 2007; Nozawa et al. 2007). Finally, AGN/QSO winds have been proposed as alternative, possible sources of dust (Elvis et al. 2002), although the expected yield is highly uncertain.

*FIRI* will allow us to trace the evolution of dust across cosmic time, in detail, and to investigate the different dust production mechanisms. Specifically:

- The sensitivity of *FIRI* will allow us to trace dust not only in rare hyperluminous objects, but also in Milky Way-like systems out to very high redshifts ($z > 6$). In particular, *FIRI* will bridge the wavelength gap between ALMA and *JWST* – crucial to determine dust mass and temperature. It will thus be possible to trace the evolution of dust mass as a function of redshift, and for different galaxy masses/types.
- Detailed sampling of the IR SED will constrain the dust emissivity, while rest-frame MIR spectra will provide information on the nature of dust through the detection of features such as PAHs and silicate emission/absorption. When compared with different models of dust formation these detailed studies will constrain the mechanism for dust production out to the crucial early stages of galaxy evolution at $z > 6$.
- The angular resolution of *FIRI* will enable us to directly locate the source of dust in distant systems. For distant QSO and AGN host galaxies, an angular resolution of ~0.02 arcsec makes it possible to distinguish whether the bulk of the dust is located in the vicinity of the QSO (< 100 pc) or in the star-forming regions of the host galaxies (i.e. SNe).

2.8 Origin, assembly and evolution of a Milky Way type galaxy

A great deal of progress has been made in the past decade on the bulk statistical properties of galaxies, including most notably the history of cosmic star formation. This beguiling and simple description of galaxy evolution has had an enormous impact, though it conceals the complexities of the merger histories of galaxies, of the mass-dependence of specific star-formation rate, of the variations in star-formation surface density within galaxies, and of the evolving relationship between the galaxy density field and the star-formation density field. *FIRI* will build on previous work in the FIR that has characterised the bulk statistical properties of the most luminous systems by allowing us to spatially resolve the physics of the star-formation processes in high-redshift galaxies.

It likely that the bulk of the stars in the disk of the Milky Way were formed or contributed through the mergers of smaller, satellite galaxies (e.g. Abadi et al. 2003). At $z \sim 1$, major mergers do not appear to be the main trigger for star formation in luminous (~$10^{11}$ $L_\odot$) IR galaxies (Elbaz et al. 2007): their rest-frame optical morphologies are predominantly spirals, with only a third showing evidence for a major merger. Furthermore, there appears to be little difference between the specific star-formation rates (SFR/M$_*$) of major merger systems and of more morphologically-quiescent IR-luminous systems. If major mergers are not the predominant trigger for star formation in these galaxies, what is? A phase of frequent accretion of gas and satellites, with concomitant vigorous star formation and vertical mixing, would be consistent with the observed distinct and homogenous nature of present-day thick disks (e.g. Seth et al. 2005); however, the links with observed high-redshift thick disks seen in the rest-frame optical are unclear (e.g. Elmegreen & Elmegreen 2006).

The coarse angular resolution of FIR instrumentation currently makes it impossible to localise the sites of star formation in these galaxies; the angular sizes are <0.5 arcsec. Do $z > 1$ galaxies follow the Schmidt-Kennicutt relationship between star formation and gas surface densities? What are the relative mass fractions of cool dust heated by the local external interstellar radiation field, and warmer dust from giant molecular clouds heated by their internal star formation? How representative are the integrated fluxes of these galaxies of the diversity of physical conditions within their ISMs? Does the external environment of a galaxy at $z > 1$ affect its internal star-formation processes (e.g. Elbaz et al. 2007)? Clearly, high-angular-resolution mapping of the star formation and gas in these galaxies is the only means to observe the predominant modes of stellar mass assembly in Milky Way progenitors.

A redshifted M82 SED has a $S_{100\mu m}/S_{850\mu m}$ flux ratio between 0.2 and 2 over the range $2 < z < 4$, and a ratio of 24 at $z = 1$. Using this SED to scale from the $S_{850\mu m} \approx 300$ μJy stacking analysis detections of Lyman-break and *Spitzer*-selected galaxies (e.g. Peacock et al. 2000; Serjeant et al. 2004), it is immediately apparent that μJy-level sensitivities are needed to map the star formation in Milky Way progenitors.

To even *detect* such galaxies requires *FIRI*'s resolution. Galaxies around the break bolometric luminosity $L_*(z)$ generate the bulk of the cosmic FIR radiation background. These galaxies dominate the volume-averaged star-formation density at any given epoch, and our Milky Way has approximately the current value of this characteristic luminosity $L_*(0)$. Semi-analytic and phenomenological models of galaxy evolution predict that resolutions of 1 to 2 arcsec are required to detect the sources that dominate the cosmic FIR background above the confusion noise (e.g. Pearson et al. 2004; Lacey et al. 2007). An immediate corollary is that sub-arcsec resolution is required merely to detect the populations fainter than $L_*$, let alone to study the physics of the assembly of their stellar populations.

Processes in the IR-luminous ($\sim 10^{11}$ $L_\odot$) galaxy population appear to be different from those in high-redshift ultraluminous ($>10^{12}$ $L_\odot$) starbursts, for which major mergers are more common triggers, and which may represent the assembly of the most massive spheroid galaxy components. Major mergers remove angular momentum (Barnes & Hernquist 1996), which leads one to expect smaller angular sizes for a fixed mass or velocity dispersion. Claims of small relative angular sizes have been made for bright (e.g. $S_{850\mu m} > 3$ mJy) submm-selected galaxies (e.g. Bouche et al. 2007), but such claims are controversial (e.g. Smail et al. 2004). Part of the problem is that the optical morphologies may not trace the star formation. Using decimetric radio emission as a proxy for the FIR, it appears that the star formation is typically confined to more compact regions within these systems (e.g. Smail et al. 2004), and possibly also that the optical morphologies are heavily affected by dust. The presence of Lyα emission lines in some (Chapman et al. 2005) suggests a porous structure, e.g. with holes created by SNe-driven winds. *FIRI* will be the only means of directly mapping the complex structure of the hot dust and star-formation surface density in these galaxies.

### 2.9 Disentangling the cosmic history of star formation from accretion onto SMBHs

Galaxy evolution is characterized by the interplay of two main phenomena across cosmic time: accretion onto SMBHs in AGN, and star formation, often occurring in energetic bursts. These two processes jointly determine the energy budget of a galaxy throughout its evolution and an evolutionary sequence from starburst-dominated through active nuclei has been indicated (e.g. Sanders et al. 1988; Storchi-Bergmann et al. 2001). On a cosmic scale, the global evolution of accretion rate appears similar to the evolution of the star-formation rate. The growth of bulges through star formation may thus be directly linked to the growth of black holes through accretion, resulting in the tight local correlation between the mass of the stellar spheroid and the central black hole (e.g. Merrit & Ferrarese 2001).

Much of galaxy evolution from the epoch of formation to the present day is hidden by large quantities of dust, causing tens or even hundreds of magnitudes of optical extinction. Unlike optical/UV observations, rest-frame MIR–FIR spectroscopy is able to trace the physical processes even if highly obscured by dust. This heavily obscured ISM is energised by the host's star formation and its growing AGN, and the IR provides the spectral diagnostics to distinguish between and quantify the two. We can thus accomplish something only dreamt of in other wavebands – measure the separate luminosity functions of accretion and star formation as a function of cosmic time. *FIRI* will be a unique tool to study the effect of both radiative and dynamical feedback: high-resolution imaging spectroscopy at rest-frame 20 to 60 μm (*observed* FIR) allows us to distinguish between the AGN-heated and starburst-heated components and constrain the possible joint evolutionary scenarios for galaxies and QSOs (e.g. Farrah et al. 2007). Imaging spectroscopy in the FIR waveband is thus the most important observational tool required to measure star formation as a function of redshift, whilst disentangling the effects of black-hole accretion, thus elucidating what we know of galaxy/QSO evolution at $z \sim 5$ to 10, the epoch of the first galaxies and SMBHs.

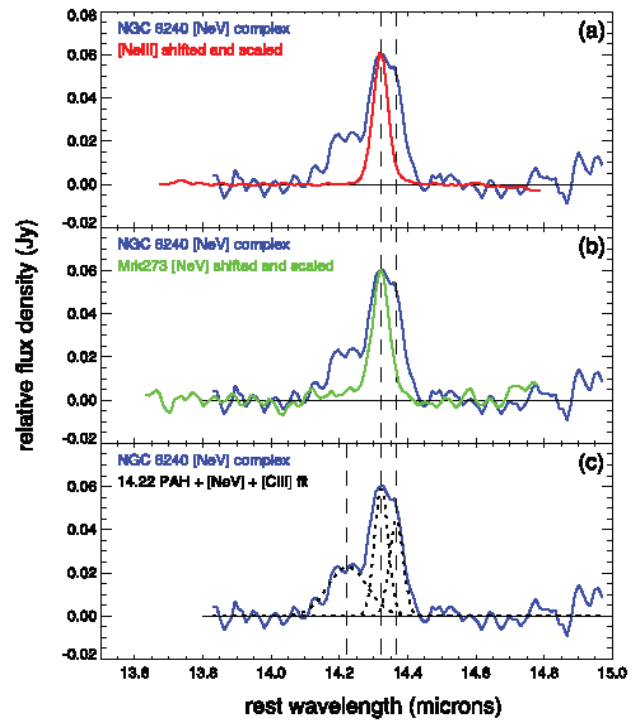

**Figure 12:** *Spitzer*-IRS observations of NGC 6240 (Armus et al. 2006). The signatures of AGN activity are apparent in this Compton-thick source from the identification of [Ne v]λ14.3μm. These lines are redshifted to > 40 μm for $z > 2$, i.e. beyond the coverage of *JWST*, but far short of ALMA's wavebands. The luminosity distance of NGC 6240 is ~100 Mpc, two orders of magnitude closer than $z \sim 2$ AGNs, indicating the need for a sensitive FIR imaging spectrometer to identify Compton-thick AGN in the distant Universe.

The rest-frame MIR and FIR wavebands contain a number of fine-structure ionic transitions from high-excitation/-ionization, non-thermally-excited, gas as well as transitions excited by starlight (e.g. Spinoglio & Malkan 1992). The FIR also includes important molecular transitions that enable us to explore the presence of dense, obscuring tori in AGN, needed to reconcile the type-1/type-2 dichotomy. The detection of OH in the local Seyfert, NGC 1068, might be due to the presence of a torus, for example (Spinoglio et al. 2005). The rest-frame MIR spectra of galaxies show aromatic PAH emission features, continua and absorption features that have important diagnostic power and are relatively easily measured since the MIR carries a significant part of a galaxy's luminosity.

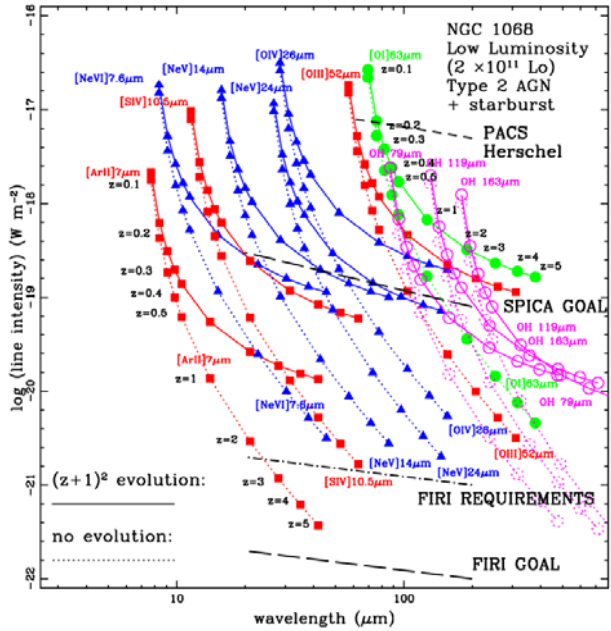

**Figure 13:** Predicted fine-structure line intensities as a function of $z$. The intensities have been scaled from the lines detected by *ISO* in NGC 1068 (Lutz et al. 2000; Spinoglio et al. 2005). Line fluxes are given in W m$^{-2}$, assuming that line luminosities scale as the bolometric luminosity and choosing two cases: *solid lines* – luminosity evolution $\propto (z + 1)^2$, consistent with *Spitzer* up to $z = 2$ (e.g. Pérez-Gonzalez et al. 2005); *dashed lines* – no luminosity evolution. *Red squares* – lines excited by stellar ionization; *blue triangles* – a AGN-excited lines; *green circles* – lines from photodissociation regions; *magenta open circles* – molecular lines from OH, possibly originating in molecular tori.

The relevant methods are well developed from *ISO* and *Spitzer* results (e.g. Genzel et al. 1998; Dale et al. 2006; Spoon et al. 2007) and have been carried to $z \sim 3$ by *Spitzer* in favourable (luminous) cases. AGN are identified in the rest-frame MIR through high-excitation NLR lines and by hot dust re-radiating absorbed primary AGN emission (Figure 12). This is a key advantage since both methods can detect Compton-thick sources that largely escape classical X-ray surveys. *Spitzer* has also demonstrated how IR features can be employed to determine the redshifts of galaxies obscured at shorter wavelengths, right out to $z \sim 3$ (e.g. Houck et al. 2005). Sensitive *FIRI* spectroscopy at low ($R \sim 100$) and moderate ($R \sim 5000$) spectral resolution will thus be the key tool to measure both star-forming and AGN components in distant galaxies as a function of cosmic time out to $z \sim 10$, over a representative range of luminosities, metallicities, and stellar masses. Many of the important features and the most important ionic fine-structure lines, capable of tracing a large range in gas density and ionization/excitation, lie in the 25- to 200-µm wavelength range for redshifted galaxies.

Figure 13 shows the expected line intensities as a function of redshift of some of these lines, assuming the cases of strong luminosity evolution [$\propto (z + 1)^2$] and no evolution, using a simple scaling of NGC 1068, a local Seyfert galaxy with both an active nucleus and strong starburst emission. The required spectral resolution is a few $\times 10^3$, and the sensitivity must be of the order $10^{-21}$ W m$^{-2}$. The spatial scales to be mapped require an angular resolution of ~0.02 arcsec, with an ~arcmin$^2$ field of view, if one wants to separate the AGN-excited gas (narrow-line regions) from the molecular tori and circum-nuclear gas excited by starlight. The angular size of a region is shown as a function of redshift in Figure 14. A FIR interferometer with spectroscopic capabilities is required to answer these questions.

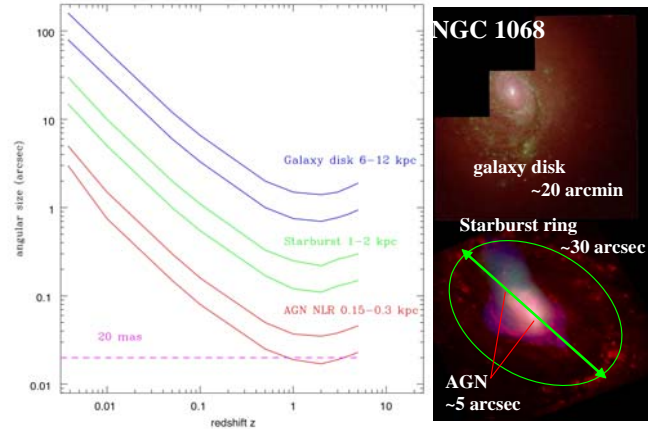

**Figure 14:** Angular resolution as a function of redshift needed to resolve: an AGN narrow-line region of 150–300 pc; a circumnuclear starburst of 1-2 kpc; a galactic disk of 6-12 kpc (for $H_0 = 71$ km s$^{-1}$ Mpc$^{-1}$, $\Omega_m = 0.27$, $\Omega_{vac} = 0.73$). On the right the three spatial scales are shown for the case of the local Seyfert, NGC 1068.

## 2.10 The obscured growth of black holes.

Once considered rare and exotic phenomena, AGN are now thought to play a crucial role in the formation and evolution of galaxies (e.g. Bower et al. 2006; Croton et al. 2006). As already mentioned, key observational support for this view comes from the finding that every nearby massive galaxy harbours a central SMBH, typically >$10^6$ M$_\odot$, with a mass directly proportional to that of its spheroid (Magorrian et al. 1998; Gebhardt et al. 2000). This seminal discovery indicates that all massive galaxies have hosted AGN activity at some point in the past ~13 Gyr, during which time the SMBH has grown, and suggests galaxies and their SMBHs grew concordantly despite vast differences in size scale. Many theoretical models suggest that joint SMBH–spheroid growth is triggered by major mergers – catalysts for galaxy-wide star formation (due to the heating and shocking of gas), driving gas towards the central SMBH. In such mergers, the SMBH can also grow by the in-spiral and subsequent coalescence of the SMBHs of the merging galaxies. However, other models show that minor mergers and galaxy bar instabilities can also efficiently fuel accretion onto the SMBH. These hypotheses can be tested with high-resolution observations of distant AGN and star-forming galaxies, probing ~100 pc scales out to $z \sim 10$. However, copious gas and dust in the vicinity of the SMBH often makes the AGN heavily obscured and challenging to identify, necessitating the need for observations at wavelengths that provide a sensitive probe of AGN activity even in the presence of absorption: the FIR.

The most "complete" census of AGN activity to date has been made from deep hard X-ray surveys, which are able to probe SMBH growth through quite large column densities of obscuring material ($N_H < 10^{24}$ cm$^{-2}$, see Brandt & Hasinger 2005 for a review). However, the absorption and scattering of photons in Compton-thick AGN ($N_H > 1.5 \times 10^{24}$ cm$^{-2}$) severely restricts the detection and identification of these sources even at hard X-ray energies, compromising our understanding of the evolution of AGN activity and the growth of SMBHs. However, Compton-thick AGNs are luminous sources of mid-IR emission, because the power that is absorbed heats the dust in the obscuring medium and is re-radiated at the local temperature. Thus even the most obscured AGN with column densities exceeding $10^{25}$ cm$^{-2}$ cannot hide in the IR.

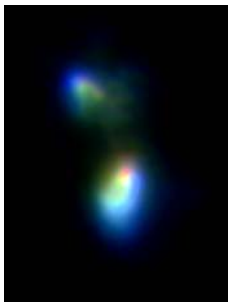

**Figure 15:** False-colour F110W-*H*-*K* image of NGC 6240, using data from both NICMOS and Keck AO, showing the two nuclei separated by ~1 kpc (Max et al. 2005). The northern part of the subnucleus is the most heavily reddened. Similar quality FIR imaging will be possible with *FIRI* out to $z \sim 10$. The observations will be less sensitive to obscuration and hence will better determine the energetics of AGN and star-formation activity than data obtained in the near-IR.

Unfortunately, representative high-redshift objects are almost always well below the confusion limit of current IR facilities such as *Spitzer*. The dust heated by the AGN, and hence the bulk of its luminosity, arises in the rest-frame MIR and hence this emission can only be reliably mapped at sub-arcsec resolution by *FIRI*, with ALMA providing a complementary view of colder dust on larger-scales. *FIRI*'s angular resolution, ~0.02 arcsec, will allow us to trace ~100-pc regions at $z \sim 10$. *FIRI* thus combines the resolution and sensitivity necessary to trace AGN out to the edge of the observable Universe.

One of the key questions about the growth of AGN is the relative importance of accretion and black hole mergers to the overall growth history of SMBHs. Although black hole merging may not impact the total electromagnetic radiation contributed to the cosmic background relative to the present day black hole mass density, merging can dramatically alter the relationship between the luminosity function of AGN and the mass function of SMBHs. With its sensitivity in the FIR, and excellent spatial resolution, *FIRI* will be able to identify multiple AGN components in $z = 1$ to 10 galaxy mergers irrespective of the degree to which they are obscured by dust and gas, and so determine the fraction of binary AGN. For example, the two AGN components in NGC 6240 (Lutz et al. 2003), separated by ~1 kpc, will be easily resolved by *FIRI* out to $z \sim 10$. Thus *FIRI* will be able to trace binary SMBHs from scales of 100s of kpc down to ~100 pc, covering 3 orders of magnitude in separation and providing a robust estimate of the birth rate of compact SMBH mergers for *LISA*, with which it is hoped black hole mergers can be identified by their gravitational wave signal. Observations with *FIRI* will reveal not only the space density of binary black holes as a function of separation, but also the distribution of mass ratios.

### 2.11 Molecular hydrogen in the first galaxies

The first stars, and perhaps the first QSOs, formed in a pristine environment of H and He, almost completely devoid of metals. *FIRI* has the potential to discover the first galaxies by detecting $H_2$ cooling lines. As the first stars formed via the collapse of metal-poor gas, heavy-element lines such as CO would have been insignificant and the main cooling will have taken place via $H_2$, e.g. the 17-μm line. The first objects might be very distant ($z \sim 20$), although metal-poor population-III star formation might have occurred at lower redshifts too ($z < 10$) and it would be exciting and important to discover these galaxies for the unique constraints on the physical processes involved in galaxy formation that they will provide.

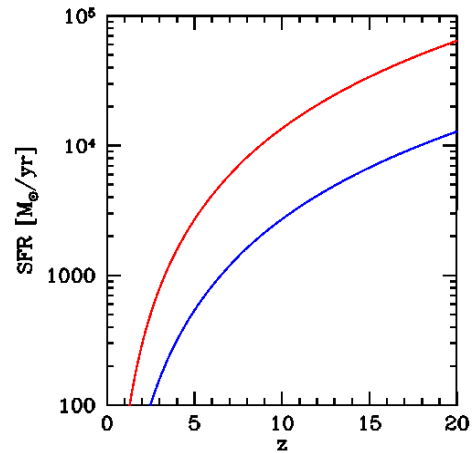

**Figure 16:** Detectability of $H_2$ cooling line emission from a gravitationally contracting metal-free cloud as a function redshift, based on the model developed by Mizusawa et al. (2005). The rate at which the cloud contracts is expressed as a SFR. The curves show the SFR at which the brightest $H_2$ cooling line is observed at a flux level of $10^{-22}$ W m$^{-2}$. The two curves show two different collapse models: one in which the collapse is spherically symmetric (red) and one in which the collapse occurs along a filament (blue).

The collapse of a metal-free cloud has been modeled by Mizusawa et al. (2005), following the chemistry, cooling and $H_2$ line emission in detail. Using the results from these authors, Figure 16 shows the detectability of the most luminous $H_2$ cooling line from a gravitationally contracting metal-free cloud at various redshifts. These models show that the cooling radiation of metal-free pre-stellar clouds in the early universe may be detectable with *FIRI*, for collapse rates of about $10^3$ $M_\odot$ yr$^{-1}$ and higher. This would correspond to the mass of a substantial galaxy cooling on a dynamical timescale. If such clouds can be detected, *FIRI* will thus probe the formation of the first luminous objects. It thus naturally takes over from both ALMA and *JWST*: where ALMA will detect metal-enriched objects, only *FIRI* can probe metal-free objects. Likewise, while *JWST* will detect objects at the epoch of re-ionisation, *FIRI* can detect the radiation directly preceding the formation of these objects.

So while the faintness of the expected signal precludes blind searches for cooling metal-free molecular clouds, there are a variety of means by which distant $H_2$ or population-III galaxies might be found. For example, objects can be pre-selected via Ly α emission using narrow-band surveys with ESO's VISTA, or using the positions of high-redshift γ-ray bursts (GRBs) from *Swift* or its successors. Spatially and kinematically resolving the line will give the distribution of the star-forming molecular gas and its dynamical mass. A typical gas-rich galaxy at $z = 10$ would, under simple assumptions (15% of total molecular gas at 165 K as in the local ULIRG, NGC 6240), produce an integrated line flux of $\sim 10^{-20}$ W m$^{-2}$ in the redshifted 17-μm line. Resolving this into 10 spatial resolution elements means detecting $\sim 10^{-21}$ W m$^{-2}$ per spatial resolution element – our requirement for *FIRI*.

## 2.12 Discovery space

We have already emphasised that observations in the FIR waveband allow us to penetrate right to the core of astrophysical phenomena, some of which we are certainly oblivious to at the current time. *FIRI* allows us to probe through the obscuring material with an angular resolution over 2 orders of magnitude better than any previous FIR observational capability, meaning that *FIRI*'s "discovery potential" is extremely high. The biggest surprise would be if there were no surprises.

## 2.13 Observatory-type science

Previous sections highlight *FIRI*'s most compelling and unique science drivers. We envisage *FIRI* as an observatory-type mission bacause of the enormous range of topics that it can address, a flavour of which we describe in what follows.

### Probing infall in star-forming regions

The water molecule is an excellent tracer of protostellar infall, much better than molecules that can be observed from the ground such as CO or HCO$^+$. *SWAS* and *Odin* showed that warm (> 300 K) interstellar regions are very rich in water, such that $H_2O$ becomes a major gas coolant. The high optical depths of the $H_2O$ lines makes them very efficient absorbers. The lines are located in the FIR part of the spectrum where warm dust provides a strong continuum background for the $H_2O$ absorption lines. While ground-based observations of CO can probe the kinematics of the gas on large scales where velocities are low, *FIRI* observations of $H_2O$ will be a unique probe of gas infall close to newly born stars.

### Probing the origin of stellar winds and outflows

One of the big scientific questions in the field of star formation concerns the interplay between disk, star and outflow. Nowadays there is agreement that jets are likely formed by the interplay of magnetic fields with ionizing winds from either the disk or stellar surface. These jets drive the molecular outflows found in every region of recent star formation. While the large-scale phenomena like outflows can be well characterized by single-dish ground-based telescopes, as well as current and future interferometers, the interaction zone – where the jets sweep up the molecular gas from their parent envelopes – will remain invisible. Fine-structure lines of ionised atoms (O III, N II, C II) are excellent tracers of these regions, suffering little extinction while being easily modelled (close to LTE, unlike recombination lines). With these lines the rotation rates of the proto-stellar jets close to their bases can be probed accurately. This information is crucial for understanding the time evolution of pronounced mass infall/accretion and mass-loss periods and thus on the final evolution of the newly formed star/extra-solar-planetary systems.

Outflows also play a large role in clearing away gas in the parent molecular clouds, effectively stopping star formation when cores lose gas. The outflow themselves are known to exhibit small scale structures associated with bow shocks when the outflow jet meets with the surrounding molecular gas. The FIR lines of OH and $H_2O$ provide the necessary tools to study these interaction zones. This is especially true when shocks occur and copious amounts of water arise behind the shock-fronts. Studies by Cabrit, Lesaffre et al. show a small number of $H_2O$ and CO lines are sufficient to accurately determine the shock properties.

### Dynamically active regions in AGB shells and young Planetary Nebulae

The spectacular evolution of circumstellar envelopes (CSE) is dominated by stellar ejections, which also play a major role in enriching the ISM in heavy elements – i.e. in the cosmic cycle. However, the basic processes (acceleration in the inner AGB circumstellar envelopes, shocks in young planetary nebulae, disks around post-AGB stars) cannot be well studied from the ground because this warm gas is not accessible via mm lines and yet not hot enough to be observed in the optical. Due to the small size of these components (< 1 arcsec), their observation requires very high angular resolution; a high spectral resolution is also needed to measure velocities, since dynamics is a crucial property.

AGB stars are evolved giants characterized by their copious ejection of mass; most stars, including our Sun, must undergo this (or a similar) evolutionary phase. The mass-loss rate is so high that it dominates the stellar evolution in this phase, in particular the transition to planetary nebulae (PNe). The evolution from an AGB star, surrounded by a CSE, to a blue dwarf in the center of a PN is the most spectacular phase in the life of most stars. In only about 1000 yr, the red giant becomes a tiny, hot blue dwarf due to the ejection of the shells surrounding the stellar core, which now becomes visible. The nebula also changes strongly during this phase. The AGB CSE, originally spherical and expanding at moderate velocity, becomes a young PN, usually showing a conspicuous axial symmetry and high velocities along the axis. A nebular component as massive as several tenths of a solar mass is accelerated to ~100 km s$^{-1}$ in < 200 yr. The mechanism responsible for this impressive evolution remains largely unknown. The interaction between very fast, collimated post-AGB jets with the fossil AGB envelope could drive this phenomenon in many cases. The most popular mechanism to explain such energetic jets involves magnetocentrifugal launching of the jet, releasing angular momentum from an accreting, rotating disk around the star or companion, in a similar way as in forming stars.

These processes are also important since mass loss by intermediate-mass stars dominates the amount of mass recycled by stars, and it is known to be the main source of ISM dust. The ejected material, merged with the ISM, is later incorporated into new generations of stars and planets, playing a major role in the cosmic cycle.

The acceleration region in AGB CSEs extends about $10^{13}$ m, or 0.2 arcsec at a typical distance of 300 pc; the subsequent cooling region is about 10× larger. The shock fronts in PNe have not been identified by Plateau de Bure Interferometer imaging so they must be smaller than 0.5 arcsec, within shell walls about 1- to 2-arcsec thick. The inner rotating disks in most objects are thought to be also about $10^{13}$-m wide (from measured IR excesses), about 0.1 arcsec at 1 kpc; only in exceptional cases (like the Red Rectangle) do they extend more more than 1 arcsec.

*FIRI* will therefore be able to map compact regions with sufficiently high spatial and spectral resolution. *FIRI* observations of AGB CSEs and PNe will thus address the main dynamical processes driving the mass ejection and wind interaction in these phases, i.e. the processes driving the spectacular stellar evolution and their crucial enrichment of the ISM with recycled material.

ISM: dynamics of photon-dominated interfaces

In the diffuse ISM the main cooling lines are the ground-state transitions of ionised C and atomic O. These lines are very bright in photon-dominated regions (PDRs), where molecular gas is being photodissociated by intense far-UV radiation. The chemistry has been progressively better understood but information on the gas dynamics requires both high spatial and spectral resolution because of the steep gradients of the physical and chemical properties. The feeding of turbulence associated with the propagation of the H II regions into molecular gas – and the role of the additional energy source provided by the dissipation of turbulence – are open questions which call for both high spectral (~0.5 km s$^{-1}$) and angular (< 1 arcsec) resolution to be able to resolve the steep gradients of physical conditions. As they are the main cooling lines, C II and O I are unique tracers for this topic. Ground-based interferometers will probe the molecular gas using rotational lines of CO (and isotopologues) and fine-structure lines of neutral carbon, providing a complete view of the structure and thermal balance.

The evolution in size of star-forming galaxies

By comparing the evolution of the FIR size of star-forming galaxies with the evolution at rest-frame optical/NIR wavelengths, the competing contributions of cosmic downsizing and hierarchical growth can be investigated and disentangled. The *Luminosity-Temperature-Radius* (*L-T-R*) relation seen locally (Chanial et al. 2007) shows how higher IR surface brightness translates into higher effective dust temperature and consequently into shifting the peak of the IR emission to shorter wavelengths. Knowledge of the size evolution makes it possible to extend tests of the local *L-T-R* relation to higher redshifts.

What angular resolution is required to resolve most high-redshift galaxies? Conservatively, we can assume that a galaxy's size can be reliably estimated, after deconvolution, for angular sizes greater than the instrument resolution. Figure 17 shows the resolving capabilities of an interferometer working at 0.02-arcsec angular resolution, with the dashed area corresponding to unresolved galaxies.

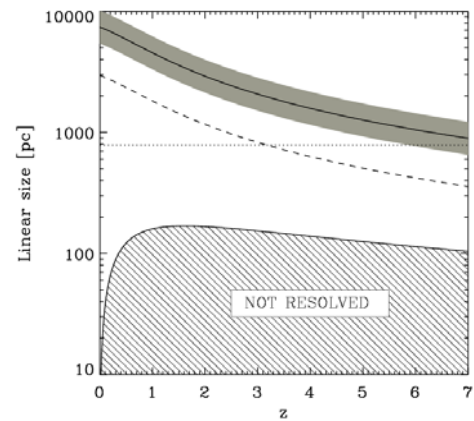

**Figure 17:** Resolving capability of *FIRI* working at 0.02-arcsec resolution. Upper solid line: empirical fiducial size evolution at rest-frame UV. 90% of IR-bright galaxies are expected to lie above the dashed line. Dotted: NGC 1569.

Determinations of galaxy size evolution in the rest-frame UV have been performed by Ferguson et al. (2004), Bouwens et al. (2004) and Papovich et al. (2005). The shaded area in Figure 17 represents the allowed range of mean size evolution based on these studies. By cross-correlating the FIRST and PSC$z$ catalogs, Chanial & Dwek (in prep) show that in the local Universe, the star-forming extent of 90% of bright IR galaxies is greater than 0.4× the mean size. Assuming that such a fraction is constant with redshift, we expect 90% of IR-bright galaxies to lie above the dashed line in Figure 17. For comparison purposes, the dotted line in Figure 17 shows the size of the dwarf galaxy, NGC 1569, derived from the radio continuum prescription (Chanial et al. 2007).

It follows that a *FIRI* capable of imaging at 0.02-arcsec resolution will be able to resolve 90% or more of the IR-bright galaxies or proto-galaxies at any redshift.

## 3. Summary of science requirements

FIR community workshops were held in Madrid (2004), Leiden (2005) and Obergurgl (2006). The participants have quantified the requirements of 26 use cases, resulting in the science requirements summarised in Table 1. The most consistent requirement – found in nearly every Galactic use case, in many extragalactic cases and in virtually all of the most important use cases – is the need for high angular resolution, 0.02 to 1 arcsec. Sensitivity is an equally demanding driver, especially for extragalactic science. Requirements on spectral resolution span the whole range, up to several $\times 10^6$; extragalactic science and around half of the Galactic use cases are satisfied with $\lambda/\delta\lambda \sim 5000$.

We have ranked each of our 26 science cases as "must do", "should do" or "could do", influenced by the science goals of Cosmic Vision, in a simple analysis of use cases taken from our science requirements document. These rankings are presented in Table 2.

## 4. Mission Profile

4.1 From science to mission requirements

From the science requirements the following conclusions can be drawn: the mission should be able to resolve proto-planetary disks and distant galaxies down to a few AU at about 100 pc distance, or ~100 pc at $z \sim 10$, i.e. angular resolution of ~0.02 arcsec. In terms of wavelength, the mission should provide access to the 28 μm rotational transition of $H_2$ and overlap with ALMA, so 25 to 385 μm. Most of the science requires extremely high sensitivity, so adequately large and cold telescopes are a necessity, as well as next-generation detectors. A field of view rather larger than the primary beam is also desirable for virtually all the high-profile science.

To achieve the required angular resolution, long baselines are necessary, ~1 km, longer than can be achieved with rigid structures. A free-flying or tethered interferometer capable of delivering excellent image fidelity is thus the instrument of choice. Unlike the optical and radio regimes, where point sources are common, the FIR wavelength region is full of structure on every spatial scale. This has profound implications for the way data must be obtained. An interferometer works well only when good *(u,v)* coverage is achieved, with attention to the short or zero spacings to allow full image reconstruction.

In recent years, four studies of FIR/submm interferometers have been performed. These were the American *SPECS* (Harwit et al. 2007) and *SPIRIT* (Leisawitz et al. 2007) studies and the European ESA-*FIRI* (Lyngvi 2006) and *ESPRIT* (Wild et al. 2006) studies. Studies for the direct detection interferometer *SPECS* (2 cold 4 m telescopes plus central beam combiner) and the heterodyne interferometer *ESPRIT* (6 ambient antennas) show that the incoherent and coherent variants of *FIRI* can take very different shapes. Neither concept on its own is capable of fulfilling all science requirements nor is any ready for immediate implementation. The science community's demands on spectral resolution range from ~5 to $10^6$. The latter is only achievable with a heterodyne system, while greater sensitivity is available at lower spectral resolution ($<10^4$) with a direct detection system. Table 2 compares the science capabilities of ESPRIT and SPECS. A figure of merit grade is assigned to each interferometer concept for each use case, and a weighted average figure of merit is calculated for the two mission concepts. Although Table 2 suggests that a direct detection interferometer is favored, neither the SPECS nor the ESPRIT concept is mature enough at this point to know with confidence which mission concept represents the better choice of detection technology. The *SPECS* study looked briefly at the possibility of combining both techniques – direct detection and heterodyne – and found no showstopper. Thus it is possible to take the prioritised science requirements and create a straw-man mission, whilst conceding that many development steps have to be taken by ESA, space industry and science/technology institutes to prove elements of the concept and assess the possible trade-offs. This straw-man mission profile should be adapted and modified according to the results of the development phase and could evolve towards the *SPECS* or *ESPRIT* designs, or neither.

Our straw-man mission profile consists of a multi-telescope (all elements of the same size) free-flying or tethered interferometer, equipped with photometric and high spectral-resolution instrumentation. Combining the high sensitivity offered by direct detection and the high spectral resolution offered by the heterodyne technique fulfils all the science requirements in this proposal, but Table 2 shows how alternative mission concepts can be evaluated, and we shall recommend below that separate direct detection and heterodyne systems be studied as well as our hybrid straw-man concept.

Together with knowledge acquired from ALMA and *Darwin* studies, we can create a concept for *FIRI* as a starting point for assessment studies, with a list of trade options to be studied in the assessment phase.

**Table 1:** Science requirements culled from use cases gathered over the period 2004-07.

| Type | λ (μm) | Δλ,Δv, Δv | R=Δλ/λ | Area,Ω (arcsec$^2$) | ΔΩ (arcsec) | N$_{fields}$ | S$_{min}$, ΔT$_{min}$ | UV constraints | Dynamic range |
|---|---|---|---|---|---|---|---|---|---|
| Water p-planetary disks | 180, 269 | 100 km/s | 10$^6$ | 10x10 | 0.02-0.1 | 40 | 20 K | Short/zero | 100 |
| Dust protoplanetary disks | 27-150 | 40 μm | 1500 | 200x200 | <0.01 | 10 | 5 μJy | | 1000 |
| Dust debris disks | 27-150 | 40 μm | 1500 | 10x10 | <0.01 | 10 | 5 μJy | | 1000 |
| HD – total gas | 112 | 100 km/s | 6x10$^6$ | 10x10 | 0.01-0.1 | 40 | 0.1K | | 100 |
| Temp/structure in trans. Disks | 63, 145 158 | 100 km/s | 6x10$^6$ | 10x10 | 0.1 | 40 | 0.1 K | Incl compact | 100 |
| The origin of outflows | 27-300 | 40 μm | 30,000 | 1x1 | 0.05 | 20 | 50 mJy | Short spcgs | >1000 |
| Lum function in clusters | 50-300 | 20-40 μm | 3-5 | 900x900 | 0.01 | 50 | 1 mJy | | >1000 |
| Cloud core stellar content | 50-300 | 70 | 1-5 | 900x900 | <0.1 | 10 | 5 μJy | Short/zero | >1000 |
| Water tracing infall | 50-600 | Few μm | 3x10$^6$ | 20x20 | <0.1 | 40 | 0.5 K | | >100 |
| Giant planets in pp-disks | 30-150 | | 1500 | 10x10 | 0.001 | 40 | 5μJy | | >1000 |
| Solar system | 50-600 | 0.1-10 GHz | 10$^6$ | 30x30 | 0.1 | 10 | 0.1 Jy | | >100 |
| PDRs | 63,158 | 100 km/s | 10$^6$ | 10x30 | 0.1 | 10 | 1 K | | >20 |
| AGB shells/young PNe | 63-300 | 200 km/s | 10$^6$ | 5x5 | 0.01-0.1 | >15 | <0.2 Jy | | >30 |
| Turbulent ISM | 75-600 | 500 km/s | 6x10$^6$ | 30x30 | 0.1 | 10 | 0.1 K | | >100 |
| Dust in ULIRGS | 24-240 | | R<5 | 10x10 | <0.05 | 1 | 1 mJy | Strong driver | >100 |
| Gas in ULIRGS | Fine-struct | 1500 km/s | 10$^3$ | 10x10 | <0.05 | 1 | 10$^{-18}$ | | >100 |
| Merging AGN | 40-220 | | <5 | 2.5x2.5 | <0.02 | 1 | 10 μJy | Point sources | >100 |
| Gas evolution | 25-200 | 1500 km/s | ~10$^3$ | 1x1 | 0.3-1 | 1 | 2 x 10$^{-22}$ | Short | >100 |
| Dust evolution | 25-200 | ~50 μm | ~100 | 2x2 | 0.3-1 | 1 | 10 μJy | Short | >100 |
| Timing of starburst and AGN phenomena | 40-660 | | <5 | 5x5 | <0.02 | 1 | 2 μJy | | >1000 |
| Probing Compton-thick AGN | 25-300 | ~40 μm | ~300 | 5x5 | 0.5-1 | 1 | 2 x 10$^{-23}$ | | >100 |
| Line diagnostics for AGN/gals | 25-660 | | ~5000 | 20x20 | <0.05 | 1 | 5 x 10$^{-21}$ | | ~1000 |
| Probing H$_2$ in blank fields | 90-400 | | ~10$^3$ | 60x60 | ~0.1 | 1 | <10$^{-23}$ | | ~1000 |
| Probing H$_2$ at known z | 50-300 | | ~10$^3$ | 3x3 | ~0.1 | 1 | ~10$^{-21}$ | | >10 |
| Imaging first massive galaxies | 30-300 | | <5 | 10x10 | <0.05 | 1 | 10 μJy | | ~1000 |
| IMF in external galaxies | 50-200 | | <5 | 5x5 | 0.005-0.02 | 10 | 1 μJy | | ~100 |

**Table 2:** Comparison of heterodyne and direct detection capabilities: scientific rankings and resulting figures of merit.

| | Type | Priority[a] | | | Weight (1 low; 5 high) | Figure of Merit[b] | | Score by Obs Type | |
|---|---|---|---|---|---|---|---|---|---|
| | | must do | should do | could do | | ESPRIT | SPECS | ESPRIT | SPECS |
| 1 | Water protoplanetary disks | x | | | 5 | 4 | 1 | 20 | 5 |
| 2 | Dust protoplanetary disks | x | | | 5 | 1 | 4 | 5 | 20 |
| 3 | Dust debris disks | | x | | 3 | 1 | 4 | 3 | 12 |
| 4 | HD - total gas | | x | | 3 | 4 | 0 | 12 | 0 |
| 5 | Temp. structure transition disks | | x | | 3 | 4 | 2 | 12 | 6 |
| 6 | Origin of outflows | | x | | 3 | 4 | 2 | 12 | 6 |
| 7 | Lum. Function in clusters | x | | | 5 | 2 | 3 | 10 | 15 |
| 8 | Cloud core stellar content | | x | | 3 | 2 | 3 | 6 | 9 |
| 9 | Water tracing infall | | x | | 3 | 4 | 1 | 12 | 3 |
| 10 | Giant planets in pp-disks | x | | | 5 | 0 | 3 | 0 | 15 |
| 11 | Solar system | | | x | 1 | 4 | 2 | 4 | 2 |
| 12 | PDRs | | | x | 1 | 4 | 1 | 4 | 1 |
| 13 | AGB shells/young PNe | | | x | 1 | 4 | 2 | 4 | 2 |
| 14 | Turbulent ISM | | | x | 1 | 3 | 0 | 3 | 0 |
| 15 | Dust in ULIRGs | x | | | 5 | 2 | 4 | 10 | 20 |
| 16 | Gas in ULIRGs | x | | | 5 | 4 | 4 | 20 | 20 |
| 17 | Merging AGN | x | | | 5 | 0 | 4 | 0 | 20 |
| 18 | Gas evolution | x | | | 5 | 0 | 4 | 0 | 20 |
| 19 | Dust evolution | x | | | 5 | 0 | 4 | 0 | 20 |
| 20 | Timing of starburst & AGN | x | | | 5 | 0 | 4 | 0 | 20 |
| 21 | Probing Compton-thick AGN | | x | | 3 | 0 | 4 | 0 | 12 |
| 22 | Line diagnostics for AGN/galaxies | x | | | 5 | 3 | 4 | 15 | 20 |
| 23 | Probing H2 in blank fields | | | x | 1 | 0 | 3 | 0 | 3 |
| 24 | Probing H2 in objects with known z | | x | | 3 | 4 | 4 | 12 | 12 |
| 25 | Imaging the first massive galaxies | | x | | 3 | 0 | 4 | 0 | 12 |
| 26 | Stellar IMF in external galaxies | | | x | 1 | 0 | 4 | 0 | 4 |
| | Priority sum | | | | 88 | | | | |
| | Score = S(priority * figure of merit) | | | | | | | 164 | 279 |
| | Weighted average science figure of merit | | | | | | | 1.9 | 3.2 |

[a] Weight definitions: 5 = "must do"; 3 = "should do"; and 1 = "could do" science.
[b] Figure of Merit definitions: 4 = fully capable of achieving science goals; 3 = not badly compromised science goals; 2 = partially compromised science goals; 1 = severely compromised science goals; 0 = impossible.

## 4.2 FIRI concept

Our straw-man *FIRI* concept consists of three 3.5-m, actively cooled afocal off-axis telescopes and a separate beam combining instrument with scanning optical delay lines and direct detectors. Because optical path length differences of the order of centimeters can be compensated with the delay line, *FIRI* does *not* require precision formation control. The primary purpose of the delay line is to enable the temporal coherence measurements required for low- to moderate-resolution spectroscopy (up to $10^4$), as in a conventional Fourier Transform Spectrometer. A metrology system provides real-time measurements of the optical path length between the telescopes and the light combination plane. The telescopes are moved to sample many baseline lengths and orientations, enabling high quality image construction. The beam combiner includes a heterodyne spectrometer, which can intercept the combined sky signals to provide high spectral resolution.

At the start of operations, as the dishes continue to cool, data will be acquired with the heterodyne system. This allows for early science to be carried out on proto-planetary disks and allows relatively simple calibration of the metrology system. When the active cooling of focal plane and telescopes is complete, direct-detection operations will start, switching back to heterodyne operations if and when assessment of the cooling-phase data suggests this is merited.

Multi-layer sunshields and cryo-coolers should be used to cool the telescopes and the beam combining optics to 4 K, where the optical elements contribute negligibly to the photon noise of the natural astrophysical background. Cryo-coolers will supplant the need for massive, voluminous cryostats, enabling *FIRI* to be launched on a single expendable rocket. *FIRI* will point approximately anti-Sunward to allow flexible baseline sampling, and the Sun shields should be sized to provide adequate field of regard. All of the desired science observations can be made if the viewable zone in a year stretches at least 20° above and below the ecliptic plane.

## 4.3 Payload

All studies so far have, for good reasons, concentrated on imaging spectroscopy of FIR sources. An interferometer is very well suited for this: a direct-detection interferometer can easily combine signals from the sky and extract spectral information when using scanning delay lines and a Fourier Transform Spectrometer. Similarly a heterodyne interferometer naturally combines both the spectral and spatial domain. We regard the equipment necessary to implement these two techniques as the payloads: the beam combiner for the direct detection interferometer, together with LO system, mixers and correlator for the heterodyne system. More details on the payload are found in Section 5.

## 4.4 Orbit

The preferred location for *FIRI* is the Sun-Earth Lagrange point, L2, which offers deep radiative cooling and minimum interference from the Earth and Moon, requiring only a single-sided sunshield. *SPECS* studies showed that all the major science goals can be achieved with the field of regard available in a large Lissajous orbit around L2. In ESA's *FIRI* study, a trailing orbit (like *Spitzer*) was also mentioned, but discarded because of ground-station difficulties. As an ESA-NASA-CSA mission, the ground station problem may vanish, but the *FIRI* data rate will be prodigious, weighing against a drift-away orbit.

## 4.5 Launch Vehicle

An Ariane 5 ECA (or the even larger American Ares V) is capable of bringing ~6600 kg to L2. The fairing length is up to 15 m while the fairing diameter is 4.57 m. The mission, as described, needs one launcher to contain sun shields, telescopes and beam combiner.

Fuel will be needed to re-orient the interferometer and for orbit maintenance, as well as for the movements necessary to fill the *(u,v)* plane if the array elements are free flying. Thrusters with low thrust and high specific impulse will be needed. Dense coverage of the *(u,v)* plane is achievable without using fuel, potentially enabling a longer mission, if the array elements are joined by tethers (Lorenzini et al. 2006).

## 4.6 Ground Segment requirements

We expect that ESA will lead the Ground Segment. European national space agencies would also participate, as well as the agency partners, NASA and CSA. For *Herschel*, the ground segment set-up used the natural capabilities of all the partners, e.g. the scientific institutes were more involved in data analysis while ESA concentrated on flight dynamics and user support. We envisage a similar scheme for *FIRI*. The ground segment set-up should be part of partner negotiations.

## 4.7 Special requirements

*FIRI* requires the simultaneous operation of multiple satellites. We have adopted free formation-flying with small thrusters as our default method of sampling the *(u,v)* plane; however, American studies show tethers as a strong alternative (Lorenzini et al. 2006), as a tethered formation could sample the *(u.v)* plane densely. Further studies should assess the costs and risks associated with tethered formation flying and determine the best operating strategy. Cooling technology is another issue. Dedicated studies should be done with respect to cooling focal-plane instruments, as well as the dishes, telescope structures and beam combiner/FTS. Cryo-coolers, such as those developed for *JWST/*MIRI and *Planck*, will supply the required cooling power for the telescopes and optics. Sub-Kelvin (~50 mK) coolers will also be required for the detectors of the direct-detection instrument. Active coolers are preferred to cryostats because cryostats are massive and bulky, and their cryogen supply is a life-limiting resource.

Regarding cleanliness, we took the base-line values for the ISO mission i.e. the obscuration by particles on a surface should be of order 30 ppm, while the molecular cleanliness level, in ng cm$^{-2}$, is expected to be of order 20 ppm. In a further study it should be investigated whether emissivity (long wavelengths) is more critical than scattering (short wavelengths). Finally, the integration and test program will be unique to FIRI. Notional integration and test plans were developed for SPIRIT and SPECS, but a more detailed plan should be developed for *FIRI* during the Assessment Phase.

At present, many of *FIRI*'s parameters are not as well defined as will be the case with other L-class missions. Integrated system studies will be necessary to put *FIRI* at a more advanced level of technological readiness. Many of the technical challenges associated with *FIRI* are common to any likely large ESA missions, though aspects of *FIRI* are easier because the wavelengths of interest are longer.

## *5. Proposed Payload*

5.1 Direct Detection Interferometer

A direct detection "optical" interferometer can be used to obtain high-angular-resolution images and spectra simultaneously over a wide field of view with a single instrument. Such an interferometer is analogous to the VLTI. For *FIRI* we envisage an extension of the optical interferometry techniques commonly used along the direction suggested by Mariotti & Ridgway (1988), who described the possibility of combining an imaging interferometer with a Fourier transform spectrometer. A further natural extension of this "double Fourier" technique involves the substitution of a detector array for the single detector used in a traditional Michelson beam combiner, a spatial multiplexing method which enables the simultaneous observation of many contiguous primary beams to cover a wide field of view (FoV), and which is currently under development in the laboratory (Leisawitz et al. 2002; Montilla et al. 2005).

At FIR wavelengths, space-based optical interferometry is not as difficult as it may seem. *JWST*, currently in development for operation at 10× shorter wavelengths, may be the largest practical single-aperture space telescope. Its 6.6-m primary mirror is comprised of 18 segments whose light must reach a common focus. That will be accomplished with a wavefront sensing and control system (Atkinson et al. 2006), a system now considered sufficiently mature for flight application. A direct detection interferometer is the optical analogue of a pair, or multiple pairs, of *JWST* mirror segments. Because wavefront sensing and control becomes easier with increasing wavelength, a FIR interferometer will be able to adopt proven *JWST* technology.

Direct detection interferometer principles

Fourier transform spectroscopy is analogous to the method employed famously by Michelson & Morley (1887). In this case light from a source is split into two beams and recombined after a moving mirror inserts a variable time delay between one beam and the other.

The combined light can be recorded on a single detector or camera pixel. The output signal intensity, when plotted against the optical delay, makes an "interferogram." The *spectrum* of the source can be constructed from the Fourier transform of the interferogram. If the optical delay range Rλ is sampled, the resulting spectral resolution will be R (= λ/δλ). In *imaging* interferometry, *spatial* coherence (interference "fringe visibility") measurements at many interferometric baselines (telescope spacings and orientations) are Fourier transformed to produce an image. Each baseline samples a single spatial Fourier component of the brightness distribution of the target scene, a component commonly identified by its coordinates in the spatial frequency *(u,v)* plane. The Van Cittert-Zernike theorem explains that an image can be constructed without information loss if enough *(u,v)* plane positions, or baselines, are sampled. By combining these techniques, as suggested above, *FIRI* will do integral field spectroscopy. When equipped with a detector array of modest pixel count the *FIRI* field of view can readily be expanded to 1 arcmin$^2$, as illustrated in Figure 18.

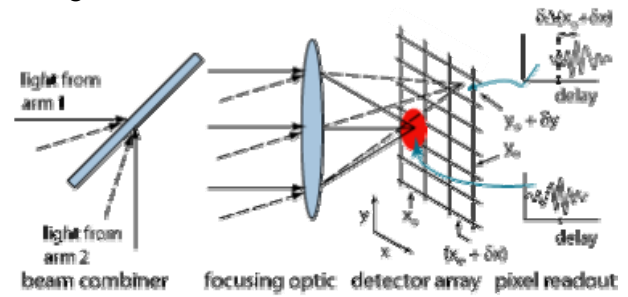

**Figure 18:** Interference fringes from field angles outside the primary beam (red ellipse) can be recorded simultaneously in the separate pixels of a detector array. If light from a source located on the optical axis of the interferometer (solid lines) is focused onto pixel $(x_o, y_o)$, then light from an off-axis source (dashed lines) might reach pixel $(x_o+\delta x, y_o+\delta y)$ after traversing opposite arms of the interferometer. The white light fringe packet in the interferogram from the latter pixel (top) is displaced relative to the interferogram from the on-axis source (bottom). The field of view is dictated by the number of pixels in the array.

Description and key characteristics

Michelson stellar interferometers can be constructed with any number of light collectors N, and the light may be non-redundantly combined pairwise in up to $N(N-1)/2$ baselines. For *FIRI*, we restrict ourselves to one of the simplest configurations, three light collectors (three baselines) and a beam combiner, all of which are on different spacecraft. The entire array rotates in a plane perpendicular to the line of sight, keeping the optical pathlength differences external to the instrument to a minimum (a few cm). The residual pathlength difference will be well within the range (~1 m) of the scanning optical delay line required both for spectroscopy and to equalise path lengths for off-axis field angles.

Guide stars are used to orient the spacecraft in absolute coordinates, and distinct sources in the science field of

view – possibly NIR point sources – serve as phase references for the optical path external to the instrument.

To synthesize as complete an aperture as possible, the baseline length B between the two telescopes can be varied continuously between ~10 m and 1 km. The minimum baseline length is determined by how close the light collectors can approach each other, and depends primarily on the sizes of the sun shields, and the architecture of the spacecraft formation. The maximum baseline is dictated by the science requirements on angular resolution (0.02 arcsec).

A direct detection interferometer has three distinct positive attributes. First, with sufficiently cold (~4 K) optical elements and low-noise detectors, its sensitivity will be limited only by statistical noise in the sky background radiation. Second, a FoV wider than the primary beam can be observed without repointing the interferometer and resampling the *(u,v)* plane. Third, the interferoemeter can provide uninterrupted access to a wide range of wavelengths. Because the detectors typically operate well over a single octave, *FIRI* will use four different detector arrays to cover the range from 25 to 385 μm. Potentially, filters could be used to narrow the bandwidth and limit background photon noise when the source of interest is seen against a bright background. A single mechanism can be used to provide the optical delay scan for all four spectral channels.

Optical delay lines

Cryogenic optical delay lines been used in space before, notably in the *Cosmic Background Explorer* FIRAS instrument and the *Casini* CIRS instrument. The FIRAS mechanism is comparable in many respects (operating temperature, lifetime requirement, physical size, and rate of motion) to the delay line mechanism needed for *FIRI*. *Darwin* has similar requirements.

Beam combiners

The beam-combiner is the heart of the interferometer system. FIR beams are more easily combined than mid-IR or shorter wavelength beams because of the relaxed tolerances on wavefront flatness. Metal mesh filters with customized spectral response functions are well suited to serve as beam combiners and dichroics to divide the broad FIRI wavelength range into octaves specified to permit an optimal match to the detectors.

The beam combining instrument grows geometrically in complexity when additional light collectors are added to a direct detection interferometer, as an additional delay line and two more detector arrays, one for each Michelson output port, are needed for each interferometric baseline. A three-telescope system should be feasible if the telescopes and beam combiner are launched together. To minimize complexity, a two-telescope system was considered in the *SPECS* study. We recommend further studies of two and three-telescope architectures.

Considerable experience in the design and operation of imaging FTS systems in the FIR is now being established. *Herschel*/SPIRE and SCUBA-2 will both have an imaging FTS and will provide experience in spectrometer operation and data reduction. This will be extended by *SPICA*/ESI, for which an imaging FTS with transition edge sensor (TES) arrays is proposed.

Detectors

In order to meet the performance requirements quoted in Section 3, multi-pixel detector arrays with NEPs of ~$10^{-20}$ W Hz$^{-1/2}$ will be needed. The *FIRI* detectors will be based on superconducting sensors and will require an operating temperature of ~50 mK. Superconducting TES arrays operating at ~0.1 K are being implemented for SCUBA-2 at the James Clerk Maxwell Telescope and are planned for other ground-based submm telescopes. While these instruments are developing and demonstrating much of the fabrication and multiplexing technology appropriate for *FIRI*, they are optimised for higher backgrounds, requiring NEPs of ~$10^{-17}$ W Hz$^{-1/2}$.

However, low-background TES arrays are also being developed for the proposed ESI instrument on the Japanese *SPICA* satellite (the subject of a complementary Cosmic Vision proposal). Cardiff University, SRON, and other *SPICA*/ESI collaborators are developing TES arrays with frequency domain multiplexing, with a target NEP of $10^{-19}$ W Hz$^{-1/2}$, deemed appropriate for *SPICA*, requiring a detector technology selection in 2009. The implementation of *SPICA*/ESI will constitute a thorough development programme for TES array and systems technology, and sub-100-mK cooler running as the last element of a cryogen-free cooling system. A natural extension of the *SPICA* detector programme would be to develop detectors with NEPs of ~$10^{-20}$ W Hz$^{-1/2}$. Similar FIR detector programmes are underway in North America. There is also a great deal of overlap between the fabrication techniques, cooling needs, and basic sensitivity requirements of detectors suitable for the next generation X-ray mission (*XEUS* or its equivalent), and a post-*Planck* CMB anisotropy mission (such as *B-Pol*).

The largest *FIRI* detector arrays, for the shortest wavelength channel, will have 32×32 pixels if the primary beam is Nyquist sampled and the FoV is 1 arcmin$^2$, or 16×16 pixels if Nyquist sampling is determined unnecessary. Smaller array dimensions cover the FoV in longer wavelength channels because the primary beam is larger.

Cryo-coolers

CEA-SBT is currently carrying out a strategic development programme for sub-100-mK cooler systems for future space science experiments, including *XEUS* and *SPICA*, and we expect these will be considered a mature technology in the next decade. A hybrid $^3$He sorption cooler/Adiabatic Demagnetisation Refrigerator is under development for *SPICA*/ESI, which can provide continuous cooling at or below 75 mK, and NASA Goddard has already achieved cooling to 30 mK with a Continuous ADR (DiPirro & Shirron 2004).

Telescope cooling to 4 K will demand a somewhat more powerful cryo-cooler than the cooler developed for *JWST*/MIRI. The *JWST* cryo-cooler was recently declared to have matured to TRL 6, and it will be adopted for flight. Experienced cryogenic engineers say that the development path from the *JWST* cooler to a cooler powerful enough for *FIRI* is straightforward.

Operation of the interferometer

A *FIRI* observation sequence comprises a slew to the target field, target acquisition (including lock on angle and zero path difference tracking), and science data acquisition. The telescopes can move during the delay line scan as long as their positions are known accurately. Data analysis will be simpler if the delay line is scanned faster than the time it takes for a telescope to move a distance equal to its diameter. The detectors will be calibrated at regular intervals. The measurements continue until all of the desired baselines are sampled.

A direct detection interferometer has a great deal of flexibility in its operation, enabling the *FIRI* to satisfy optimally a variety of measurement requirements. For example, if a particular observation calls for angular resolution coarser than 0.02 arcsec, spectral resolution <3000, or a FoV smaller than 1 arcmin$^2$, then only short baselines, a restricted delay line scan range, or a smaller number of pixels can be read out, respectively.

Thermal sources similar to the micro-lamps used in *Spitzer*/IRAC are sufficient for calibration. These lamps should ideally be installed near a pupil stop or an image of the pupil. The lamps will need to be sized to produce a reasonable power per detector (for instance, 500aW, 5fW, and 50fW), tailored for each band. The illuminators should be stable and fast ($t < 1$ s). To calibrate data, the illuminators would be cycled on and off as necessary after every delay line stroke, while the mirrors are moving. This illumination response would be used to transfer calibrations from standards to unknown sources. Calibration can take place in about 5 s, much less than the time to move a mirror.

Performance with respect to the science requirements

A direct detection interferometer satisfies all of the extragalactic science use cases in the *FIRI* Science Requirements Document, as well as all of the Galactic use cases asking for low to intermediate spectral resolution, as indicated in Table 1.

Current heritage and Technology Readiness Level

J-T and Stirling coolers for telescopes and instrument cooling: based on upgrades of flight-qualified coolers. Laboratory tests have demonstrated heat lift capacity exceeding requirements. MIRI cryo-cooler: TRL 6.

Sub-K cooler: Development activity within Europe to develop ADR coolers for *XEUS* and *SPICA*, and the Goddard Continuous ADR: TRL 4.

TES Detectors: Europe: SCUBA-2 (being built) and *SPICA*/ESI (in development). TES systems are in development in North America for ground-based and space-borne applications (e.g., JPL, Berkeley, NASA-GSFC) Additional development will be needed to achieve lower NEP: TRL 4.

Imaging FTS: SCUBA-2 and *Herschel* will operate imaging FTS instruments in the 2008 time frame. A low-background space-borne imaging FTS is being studied and developed for *SPICA*/ESI: TRL 6.

The Wide-field Imaging Interferometry Testbed at NASA Goddard collects data representative of those obtainable with *FIRI* (only instrumental effects contribute to the phase noise) and will address three unsolved problems: (a) For astronomically interesting scenes, what combinations of photon noise, spatial frequency undersampling, and motion smearing during the observation are tolerable?, (b) How good will *FIRI's* spatial-spectral images be?, and (c) How sensitive can we expect a direct detection FIR interferometer to be, and what factors will limit the achievable sensitivity?

Procurement approach and international partners

Science and technology institutes in Europe and North America are well equipped to provide the crucial ingredients of a direct detection system. These include the optical components, detectors, sub-K cooler, FTS and associated warm electronics. ESA and NASA can take care of parts procurement. A European company could be prime contractor for the telescopes. Many of the beam combiner components and the metrology system can be provided by the space industry, while special components, such as the beam-splitters, filters, and detectors could come from technology institutes. The beam-combining instrument can be assembled and flight qualified at an institute experienced in the development of science instruments for space telescopes, such as NASA's GSFC.

Assembly, Integration and Verification

The assembly, integration and verification (AIV) process for *FIRI* should begin with high-fidelity software simulations and test-beds to validate designs, and should lead to the development of proto-flight hardware. Unique cryogenic test facilities (similar to that proposed for *SPICA*) owned by ESA and NASA can be used to test the individual *FIRI* telescopes and the beam combiner. Performance testing of the entire interferometer may be impossible or impractical, but verification of the beam combiner's performance will suffice if appropriate tests are conducted, if the telescopes are separately proven to provide collimated FIR beams, and if tests are conducted to ensure that stray thermal radiation in excess of the design tolerance will not be able to reach the detectors.

Critical issues

Although the direct detector interferometer has a lot of heritage from *Herschel*, and probably from ESI on *SPICA*, there are several critical issues:
- Detector array development to achieve very low NEPs and high read-out rates;
- Thermal model validation and verification

- What is the best optical design and how do we do stray-light suppression?
- Beam combiner engineering design
- High specific impulse propulsion system
- Power system to drive thrusters, operate mechanisms, detectors, cryo-coolers, communication system, etc.
- How best to actuate mechanisms intended to work at 4 K with minimal parasitic heat load?
- Sky density and quality of phase reference sources in the field of view?
- What is the best AIV approach?

5.2 Heterodyne Interferometer

The heterodyne interferometer will be the instrument of choice when high spectral resolution is required. It provides this resolution naturally in the mixing and down-converting process; at the same time, unlimited amplification and splitting of the down-converted (Intermediate Frequency, IF) signal is possible. As such it has been the technology of choice for ALMA.

A description of the principle is given below, followed by a more detailed description of key elements: mixers and local oscillators; correlators; and observing modes, metrology and $(u,v)$ plane filling.

Heterodyne interferometer principles

In a heterodyne system the sky signal is combined with an internally generated Local Oscillator (LO) signal. When this combination is done in a non-linear mixing element, the result is a copy of the sky signal but amplified and down-converted from the THz range into the GHz range. In this regime, low-noise amplifiers are available to further amplify the signal. At this stage the signal is electrical rather than optical. The amplified signal can be split when needed and correlated with other "sky" (or better, amplified IF) signals in digital correlators. The only actively cooled parts in the spacecraft at this time are the detectors (mixers). The high angular resolution guarantees that the telescope background is not an issue. An ambient telescope, like *Herschel*, suffices at these early stages.

There are several ways of correlating the IF signals. The ALMA case is based on a XF correlation scheme, while the advantage of novel FFT Spectrometers is that they can operate in a FX correlation scheme. For *FIRI*, a hybrid correlator (FX then XF) may well be the best option. A big advantage of a heterodyne interferometer is that the very high spectral resolution guarantees a very high correlation length; delays can thus be applied electronically or in software based on a geometrical model. This feature makes it possible to keep the metrology of a multi-telescope system rather relaxed, i.e. observing is possible while the telescopes are moving. In fact, correlation is possible with a prediction of the position of the spacecraft within tens of cm while the metrology of spacecraft position with respect to each other should be a few μm to guarantee good phase calibration. The heterodyne concept could make use of a central correlator, like the direct-detection beam combiner, but a distributed correlator (in all available spacecraft) is also possible and probably preferred.

Since moving the telescopes while observing is not a problem, $(u,v)$ plane sampling can be achieved by letting the dishes move outwards (after acceleration), sampling during this floating phase. Deceleration to a stop and accelaration inwards is necessary to bring the array back to its initial configuration, after which the array can move to another source. Since delays can be done electronically or in software, the array could even fly in a 3-D configuration, thereby mitigating collision risks.

The heterodyne interferometer thus consists of several telescope spacecraft with, in their focal planes, several heterodyne mixers tunable at spot frequencies in the FIR. Mixers receive their LO signal from dedicated Local Oscillators, with phases tuned to a master LO phase distributor which distributes phases using a detailed geometrical model. These signals are down-converted in the mixers and amplified several times. The IF signals (which are led to the service module of each spacecraft) can be split and distributed to the other spacecraft for correlation, where delays will applied in electronics and software, without moving optical parts. An alternative would be to send all IF signals to a central correlator where visibilities are calculated and sent to Earth at a low data rate. Mosaicing will be possible during each observation, increasing the field of view of the heterodyne *FIRI* complement.

Mixers and Local Oscillators

Based on knowledge of ground-based telescopes and the developments for *Herschel*/HIFI, we have identified two possible types of mixer: Hot Electron Bolometer (HEB; >1.5THz) and Superconductor-Insulator-Superconductor (SIS; <1.5THz). These mixers can be used immediately in an interferometer, but for optimal bandwidth, stability and sensitivity, extra development is needed. With HIFI, in principle the technological readiness rises to 9, while very high frequency HEBs have been demonstrated in the lab (TRL 5). Below 2 THz, HIFI shows that solid state LOs as provided by JPL (USA) and RPG (D) can be built. However, above 2 THz the solution should come from new developments, e.g. Quantum Cascade Lasers. These devices need development to decrease their power consumption and improve phase locking. If these developments lead to improved QCLs, these are promising devices at THz frequencies. With HIFI, the TRL of the solid state LOs is 9. The QCL TRL is 4.

Correlation

There is a choice of two different correlation schemes (multiplication and Fourier transform) generally known as XF, FX or hybrid schemes. All these schemes are planned to work on ground-based telescopes (e.g. ALMA: XF; SKA: FX) or work already (e.g. APEX: FX). The current development of FPGA's is very fast and it is clear that cross-correlation will not be an issue for *FIRI*. There is, however, a need to do a thorough trade-off study between the different correlator concepts. These studies should include the correlation schemes, the IF treatment (in steps of 1 GHz, or larger), and the distributed versus central correlator approach. Power, volume and mass could be better estimated than at present. The rapid eletronical revolution to FPGA-based devices makes that 10× less power in a few years is likely achievable. Also mass and volume are reduced.

Geometrical model, LO distribution and locking

In a submm or FIR heterodyne receiver, the LO that drives the mixer must operate at a fixed and well-defined frequency, with a high degree of spectral purity. In the case of a heterodyne interferometer, this is taken one step further, in that the coherent combination of the IF signals from each antenna requires that the LOs at each antenna are operating in phase. Moreover, in combining the signals from each antenna in a correlator, it is also necessary that the total optical and electrical path lengths from source to correlator are equalized for each antenna by introducing antenna-specific delays somewhere in the system. Due to the coherent nature of heterodyne detection, precise optical delays lines (which require precision optical systems and high-reliability mechanisms) can be replaced by precisely defined timing delays in the correlation of the data.

It is assumed that for *FIRI* all the reference signals originate at one central point, which can be any of the satellites. For redundancy purposes, each satellite can be provided with its own reference oscillator. This oscillator can also be used in the GPS-like coarse positioning network, which will require an independent time/phase reference at each satellite to be compared with the received signals.

Gross delay compensation should be done after digitisation and requires accurate computer control. It can either be done on-board, at the receiving element, or in the correlator element. It is probably preferable to have it where the (rest of the) fringe stopping is done.

It seems attractive to accommodate the fringe stopping entirely in the correlator by accomplishing fringe rotation after digitisation. This has the advantage that it reduces the complexity of the LO distribution and limits the number of interfaces. If, however, the large desired bandwidth reduces the number of bits in the signal representation, this approach has limited ability to track the phase, resulting in a loss of signal to noise. This could be remedied by fringe stopping in an FX scheme, or after correlation, but *FIRI* fringe rates are still high. Further study is probably needed. Thus, the classical approach to modifying the LO seems to be the appropriate solution for *FIRI*, despite the fact that it implies some extra complexity for the distribution scheme. In any case, some complexity is inevitable in order to accommodate phase switching and possible side-band rejection.

In the preferred scheme, each element has the necessary logic to evaluate the correlator model. Input to the model calculations is the detailed geometry of the array and the most demanding part for *FIRI* is that the relative location of the antennas must be known a-priori. However, as the antennas are moving on perfect linear tracks this is not a problem. Furthermore, the orientation of the array in space must be known, as must the direction of the target and the motion of the array with respect to the target.

Detailed overall system design is required to validate the above approach. Although we see no fundamental problem, it is difficult to verify it without details of many different components.

Operation of the interferometer

Heterodyne interferometers can boast of their large correlation length. WSRT, VLA and ALMA benefit from the Earth's rotation, which ensures that the projections of baselines on the sky change continuously, thus filling the *(u,v)* plane, necessary to create high-quality images. However, there is no daily rotation that can be used at L2, so filling the *(u,v)* plane must be accomplished in a different way. An advantage of the *ESPRIT* design stems from its coherence length. Due to its very high spectral resolution, about 2 MHz, this length is about 150 m (for continuum with 4 GHz resolution, this reduces to 75 mm). Within this distance wave fronts can be considered to be coherent in nature, and thus when signals from two telescopes need to be correlated their path length difference needs to be known within this coherence length and correlation can take place using electrical delay lines. This allows for observing while the telescopes are moving and thus the *(u,v)* plane can be filled "on-the-fly". For single spectral lines, the advantage is even greater.

In general, asymmetric configurations like those used for LOFAR, ALMA or the VLA are preferable. An interferometer in space can benefit from the possibility of using a continuously changing 3-D configuration; a study should look at the possibility of adding elements.

An important aspect is the geometrical LO distribution model, as defined above. This model needs, as inputs, the absolute velocities and positions of each element's phase center in order to calculate clock adjustments for the correlator and the phase differences that are sent to each LO. In doing so, one ensures only the geometrical delay must be dealt with by the correlator.

Determining interferometer configurations is not easy. No standard has ever been established for ground-based interferometry, let alone for space. *Darwin* studies have been conducted, but these concentrated on symmetric configurations which are less suitable for a heterodyne interferometer. It is therefore necessary to study optimal configurations. This should take into account the following restrictions: 1) each "observing run" should result in a good image; thus, the *(u,v)* plane needs to be properly sampled; 2) it should be possible to point the telescopes in the same direction within 1/20th of the primary beam size; 3) it should be possible to keep the metrology system working (i.e. each telescope needs to see the other telescopes' metrology receivers); 4) it should take into account the acceleration the array gets from its thrusters, and it should take into account any velocity within the configuration.

Performance with respect to science objectives

At 100 μm, with a velocity resolution of 1 km s$^{-1}$ and a angular resolution of 0.02 arcsec, in one day, a line sensitivity of 18 K can be obtained for mixers with a 1000-K system temperature (2× better than HIFI). Continuum sensitivity is 3 mJy beam$^{-1}$ (r.m.s.). Since almost all radiation emitted in the FIR is thermal emission coming from regions of 30 K and higher, it is clear that every optically thick line coming from 0.02-arcsec areas can be seen in 1 to 2 days. In our straw-man mission profile we have used 3 dishes to optimise the synergy between the two detection systems, so the same cases will take longer to reach the same senistivity. However, adding dishes is an option that should be studied, because it adds to the sensitivity of the system and it provides a big leap in the speed with which the (u,v) plane is filled.

A heterodyne interferometer tackles the high-spectral-resolution science in Section 2. As such, a heterodyne instrument suite on *FIRI* is compliant with Cosmic Vision in the area of planet and star formation.

Bandpass calibration

In the FIR, most sources are extended, so the usual method of measuring gains is not easily applied. *FIRI* has to rely, in the heterodyne case, on careful regular observations of its own antenna-based calibrators, together with (periodic) observations of the astrophysical objects suitable for bandpass calibration (these will probably be the same for ALMA and *FIRI*).

An overview on bandpass calibration for submm interferometers is given by Bacmann & Guilloteau (2004). Their method is applicable to *FIRI*, when all atmospheric contributions are assumed to be equal to 0 (or atmospheric gains equal to unity). Inspecting their equations 10 and 11, we see that the amplitude bandpass is independent of time and is equal to the product of the respective gains of the receiver, IF, filters and converters. Together, these two antenna-based quantities can be used to determine the baseline-based visibility gains that are necessary for real calibration. The phase bandpass is, to first order, the delay and this geometrical effect is treated below

So, it only remains to determine the antenna-based amplitude and bandpass. This can be done in several ways for ALMA, depending on loads at each telescope. Detailed calculations should show whether this, as for HIFI, is a feasible option. For *FIRI* the bandpass calibration will probably rely on the availability of internal calibrators. Further studies into the applicability of antenna-based gains for the determination of baseline-based bandpass solutions are necessary.

Phase Calibration

Every interferometer relies on accurate knowledge of the phase, because this is where the information is that is used to create images. Therefore, the phase differences between the different telescopes have to be known accurately, which is in space only the geometrical path difference or delay, and the internal phase differences. The advantage of a heterodyne *FIRI* over ground-based radio telescopes is the lack of the Earth's atmosphere. This means that there are no phase fluctuations in front of the array. All phase errors are either geometrical, electronic (generated in the system itself), or due to jitter of the (master-) LO.

A very stringent limit on the electronic behaviour of phase is needed and geometrical errors must be as small as possible (as a rule of thumb, 0.1 rad is generally used in ground-based interferometry). Once suitable celestial sources for phase calibration are found, an algorithm is needed to tie the phases together. In principle, ALMA- or SMA-like schemes can be used where observations of phase-calibration sources in three places in the sky and separated at least by 30° yields enough information to obtain the phases of each individual telescope.

In general, *FIRI* should be much less susceptible to phase changes than a ground-based interferometer. This should offset the poor availability of phase calibrators.

For *FIRI*, the main LO phase change is from the geometrical delay. A well-designed metrology system will provide exact delays that can be fed into the correlator and sent out to change the slave LO phases with respect to the "master" LO. The second-most important change in LO phase stems from drifts in the electronics within each satellite. Thus, priority should be given to producing designs for the electronics that suppress drifts. As a second step, information about electronic phase drifts should be derived from comparisons of signals produced by the electronics units with external references. Implementation of such a technique is still to be investigated.

A geometrical model must be made that implements all known positions, velocities and accelerations, and outputs specific LO phase differences to all LOs in the system. Studies should show whether this is a viable plan. Secondly studies should be started that address the question of synchronizing the LO drifts from all elements of the satellite. Third, sources on the sky should be found in which an SMA-like scheme (observing three sources to obtain the phase of each element) can be applied. A study should be undertaken to determine an optimal calibration strategy.

Critical issues

While many aspects of the heterodyne payload can be based on HIFI heritage it is clear that much development is still needed. Amongst these we single out the following:

- Detector development: can we approach the quantum limit whilst increasing IF bandwidth?
- Can we make reliable LOs at THz frequencies?
- What is the best, most compact, power-efficient cross-correlator?
- Focal plane: how do we cool the mixers to 4 K?

- What is the best AIV approach?

## Procurement approach & international partners

Science and technology institutes in Europe and North America are well equipped to provide the ingredients of a heterodyne system: the mixing devices, LOs and correlators. The front-end optics, LO distribution and metrology can easily be provided by space industry. There is an advantage if ESA both procures and pays for the parts. AIV will be difficult and would benefit from simulators, test beds and a low-frequency channel in atmospheric windows (e.g. at 100 GHz).

Cooling for the focal-plane detectors may use the sorption micro-coolers developed for *Darwin*, which may be further developed by space industry.

## 5.3 Common Characteristics

### Pointing and alignment requirements

*FIRI* has a primary beam of 6 arcsec at 100 μm and a spatial resolution of 0.02 arcsec. In order to make sure that the electrical phase centres for the heterodyne interferometer are aligned well, the pointing should be of order 1/20th of the primary beam: 0.3 arcsec. *Spitzer* currently achieves 0.45 arcsec, so current figures have to be improved upon by only a factor ~2 in the coming years. Pointing is slightly less relevant for the direct-detection system, because of the use of a larger field of view, however in order to maintain the fringe-stop, the system would also benefit from 0.3-arcsec accuracy.

Pointing stability is generally guaranteed by reaction wheels (RWs). In order to off-load extra torsion, as with *Spitzer*, nitrogen could be blown off.

Alignment has been tackled for *Herschel* by careful planning of all mechanical interfaces, using optical light for tests. This implies all interfaces should be reflective (or in some cases transparent) to optical light. In cooling down there will be shifts going from ambient (80 K) to actively cooled (~4 K).

Alignment also influences the metrology beams, the heterodyne data beams and the optical beams used for optical correlation. The interferometer configuration (filling the *(u,v)* plane) should be such that metrology is kept during the observation, in order to preserve knowledge and control of the path-length differences.

## *6. Basic spacecraft key factors*

This proposal should be seen as the next step to achieve a *FIRI* that is based on existing, European technologies as far as feasible, but also includes new technologies in order to be able to remain within reasonable cost, mass and power budgets and to achieve the desired performance. Many of the new technologies proposed here are related to the interferometry character of the mission, and can be borrowed from the mission studies done for *Darwin* or *LISA*; NASA's interferometry missions (e.g. *TPF*) can also contribute. Technologies developed for micro-satellites by several scientific groups and companies (e.g. Surrey Space Centre, Guildford, UK, http://www.ee.surrey.ac.uk/SSC/) may lead to a light-weight bus suitable for larger missions like *FIRI*. The weight goal should be 300 kg/satellite.

### 6.1 Number of spacecraft

There are a number of reasons that lead to our current concept of three dishes plus one beam combiner for *FIRI*. It should be possible to have phase closure between the interferometer elements which leads to at least three heterodyne elements. Furthermore, there must be enough sensitivity to reach the line brightness in the warm gas, which would benefit from as many elements as possible. For the cooled *FIRI* complement, however, there is little benefit in using more than three collector spacecraft. Calculations for a three-element heterodyne interferometer with dishes of 3.5-m diameter show that this is sufficient to fulfil the sensitivity requirement for the high spectral resolution cases as long as the detectors have near-quantum limited performance. A similar sensitivity can be reached with more baselines in a shorter time when more elements are used.

The direct-detection interferometer needs a central beam-combining satellite in addition to the collector apertures. All four have to be well shielded from the heat loads of sun, earth and moon with several sun shields and baffles, making the four satellites rather heavy using today's technology. However, with clever low-weight instrument design it will be possible to pack everything in one launcher fairing. This should be confirmed by a dedicated assessment study. The 3-element direct-detection interferometer will fulfil the sensitivity requirement as long as the detectors reach NEP values close to $10^{-20}$ W $Hz^{-1/2}$, and the cooling of the whole system attains temperatures close to 4 K.

### 6.2 Attitude and orbit control

From first principles, the elements should be pointed with an accuracy of 0.3 arcsec. Focal plane arrays can live with more relaxed pointing requirements. We assume that the interferometer should be 3-axis stabilized in each element.

For thermal reasons, and because of its benign radiation environment, L2 is the orbit of choice for *FIRI*. The interferometer should be able to move its elements inwards and outwards to fill the *(u,v)* plane in a controlled fashion. This will require, e.g., small ion motors. Repointing the whole array is necessary to move to another source.

We should ensure that the number of guide stars is adequate. In the *SPECS* study it was found that the only safe way to point with sufficient accuracy was to run NIR radiation from stars through essentially all the optical components of the interferometer. To do this efficiently, this would need to be done at wavelengths of about 2 μm, so the telescopes and optics need to be of NIR surface quality and NIR cameras will be needed for guiding. This may be a problem that needs significant attention in a phase A study.

### 6.3 On-board data handling and telemetry requirements

A heterodyne interferometer collects data at high speed and in large quantities. Assuming we will have (for two polarisations) a conservative 4-GHz-wide Intermediate Frequency (IF) signal out of the mixers, we can do the following calculations. The IF signal needs to be sampled at roughly twice (Nyquist) the IF frequency in order to prevent aliasing. This signal can be digitised by a 2-bit A/D converter. There are several ways to proceed with correlation, of which one is a distributed correlator, where the IF signal is split in e.g. three or more basebands before it is sampled in the A/D converter. With the redundant properties of distributed correlators the single-point failure of a central correlator can be avoided and the data flow in the interferometer can be reduced. For a 3-element interferometer and three basebands each element transmits 10 Gb/s and receives 20 Gb/s. These data rates are high, but within each element such rates are easily handled. Outside the elements simple IR links can transmit and receive these data rates. Finally, after correlation, the visibilities (of order 100kb/s averaged over one day) can easily be transmitted with radio links as used e.g. for *Herschel*. There is a need for mass-memory on-board, for data storage, when no downlink is possible, similar in size as available to *Herschel* (~ 220 Gbits).

The raw data rate from a direct detection interferometer is approximately $b_{max}^2 \eta_{uv} \theta_{FOV}^2 n(R\lambda + b_{max} \sin\theta_{FOV})/\lambda^3 t$, where $b_{max}$ = 1 km, the longest baseline length sampled, $\eta_{uv}$ is the *(u,v)* plane filling fraction, $\theta_{FOV}$ is the field of view diameter, n (~4) is the number of samples per fringe, R is the spectral resolution, $\lambda$ is the wavelength, and t is the observation time per target field. By implementing the inteferometer design features, mode of operation and sampling compromises recommended in Appendix F of the *SPECS* study report, the data rate can be reduced from a raw rate of 2 Gb/s to ~100 Mb/s.

Since the interferometer has only one operating mode, telecommands will be very simple and mainly deal with setting up the instrument and with internal calibration procedures. Downlinked telemetry of science and housekeeping data can be handled via high gain antenna in $K_a$ band and delivered to a single ground station in approximately 30 minutes per day.

### 6.4 Mission operations concept (Ground Segment)

*FIRI* is envisaged to be an observatory with the traditional elements of a ground segment like e.g. *Herschel*. The Ground Segment will consist of the usual ground stations – a Mission Operations Centre, and a Mission Science Centre catering for User Support, Proposal Handling, Observation Planning and Data Processing, Distribution and Archiving. Orbital dynamics and pointing will be ESA responsibilities, and instrument monitoring, calibration and optimisation will be the responsibility of the instrument consortia. According to current ideas, individual *FIRI* observations will require about two days, making the whole set-up of the ground segment and in particular the operations easier. We expect that the *FIRI* instrument consortia will support the equivalent of the Herschel Instrument Control Centres (ICCs).

There will be one ground station provided by ESA, but a ground station provided by NASA is considered an option in this proposal. The ground station will provide the usual functions: uplink of telecommands, collection of the down-linked telemetry consisting of house-keeping data as well as science data, contact via a transponder for ranging and support services. The ESA-provided *FIRI* Mission Operations Centre will be responsible for first inspection and re-ordering of the data before the data are sent to the astrophysical community. Astronomers will gain access to *FIRI* through a peer-reviewed proposal process, handled by ESA. The software for the ground segment should be a joint development between ESA, NASA, CSA and the *FIRI* national agencies. They should build the architecture and tools for an easy dissemination of the data. The fact that only two instruments are present with one operation mode each implies a much easier implementation of commanding and data reduction than was the case for e.g. *Herschel*.

### 6.5 Estimated overall resources (mass and power)

Reasonable estimates have been assembled from the literature, but further assessment studies should reduce the uncertainty in the budget numbers presented below.

#### Attitude determination and control

ESA-*FIRI* describes in pages 250 to 257 "Guidance, Navigation and Control". Their design is classical: RWs, thrusters, sun sensors and rate sensors.

The large reaction wheels of ESA-*FIRI* are used to rotate the 30-m boom and that is not relevant for our *FIRI* design. Each *FIRI* satellite only needs to maintain its pointing. The classical RWs of ESA-*FIRI* could be replaced by Control Moment Gyros (CMGs) – a momentum wheel, spinning at constant speed, mounted in gimbals fitted with torque motors. The motors torque the momentum vector to change its direction. As the momentum direction is changed a reaction torque is created on the spacecraft. These are more efficient than reaction wheels, but presently complex and expensive. CMGs have been used in military-related missions. Fortunately, cheap and light-weight micro CMGs are presently developed to be used for agile small satellites and have been tested in orbit (Lappas et al. 2006).

Four wheels are needed (one for redundancy) and these four will weigh 19.4 kg. CMGs will weigh less: four Lappas CMGs will weigh 8.8 kg. As other equipment the ESA-*FIRI* document lists 3 sun acquisition sensors, 3 star trackers and their associated electronics box, 2 interial reference units, and 2 attitude anomaly detectors. All these together weigh 17.7 kg and consume 68 W. Each *FIRI* spacecraft will need this classical equipment.

The inertial reference unit will presumably contain 2 gyros. A mechanical gyro, a mass on a gimbal senses the rotation rate and therefore accumulates error. By using the star tracker regularly the gyros are calibrated. Nowadays there also exist ring laser gyros, where time around loop and speed of light are used to calculate rate. Such ring laser gyros require much less star tracker calibrations. They are used on the Boeing 757 and 767. They can be bought from a number of firms, e.g., Kearfott's KN-4070, weighing less than 5 kg. Lighter space-qualified units are developed by several firms. These units can be considered to be of TRL 7-8.

The *Darwin* system assessment study (Alcatel, 2007, page 28) says that each *Darwin* spacecraft needs 63 kg hydrazine to move it to L2 and uses 4.5 kg of propellant for manoeuvres, over 10 yr. For *FIRI* this will be similar, but more studies are needed to assess the fuel constraints. The TRL status of thusters like FEEPs is ~7; they have been demonstrated in a mission to the moon. The 63 kg hydrazine is ascribed to the launch system.

Total weight of Attitude Determination and Control sub-system could be 25 kg, while the thruster system will weigh 40 kg (Alcatel 2007), so 65 kg for the whole system. The attitude system will consume 50 W (extrapolating the conservative 68 W to the future).

Telescope and support structures

The telescope of *Herschel*, made of SiC, weighs 310 kg, too heavy for *FIRI* with three telescopes. New techniques are needed to reduce this mass to acceptable values. Tan et al. (2005) describe a 5-m diameter inflatable parabolic reflector; a model has been made and integrated. The complete weight is presently 33 kg. It has a TRL of 4. Alternatively, Composite Mirrors Associated in the USA, is capable of producing carbon-fiber mirrors of *Herschel* size weighing <100 kg.

Tromp (2005) describes making structures using a 5-axis milling machine. The trick is that each excavated component has a front- and backside stiffening the structure considerably. This may decrease the weight of many mechanical components with 50%.

The telescope and all its structures may weigh <50 kg. When all other mechanical structures are made according to the ASTRON-Tromp method this may lead to a weight allotted to telescope, mechanisms and connecting/supporting structures of 100 kg.

Power sub-system; solar panels

For *Herschel* the numbers are: at BOL (begin of life) and EOL (end of life) the *Herschel* solar array will generate > 1674 W, eventually 1475 W, using 3 panels of 10.8 m$^2$ and weighing 140 kg. Thus: 137 W m$^{-2}$ and 10.5 W/kg. The cell efficiency is 27%.

A number of studies in recent symposia show that these numbers can be improved considerably in the next 5 to 10 years. We mention a few in what follows.

Bett et al. (2005) show that efficiencies of 42% can be reached by using cells with 5 junctions. Samson et al. (2005) indicate that the weight and cost of thin film solar arrays will be 50% lower.

Brandhorst et al. (2003) describe using a 8.5-cm Si Fresnel lens to focus sunlight 8 times. The test system, SCARLET, achieved over 200 W m$^{-2}$ and over 45 W/kg, in space. A 7,129-kW, 24-m$^2$ version of this array exists, weighing 38.9 kg, i.e. 300 W m$^{-2}$ and 200 W/kg. TRL 7. Lefevre et al. (2005) presented a study on Deployable structures for flexible solar generators: 19.4 kg, 15 m$^2$ blanket, 163 litre packaging, 10% efficiency solar cells.

If solar panels for each *FIRI* satellite use the 8 times concentrator technique, an EOL of 1500 W/satellite can be obtained with 7.5 kg and 5 m$^2$ of solar panels.

Pages 215 to 225 of ESA-*FIRI* describe the power system. An electric power system consists of the components: power source, energy storage, power distribution, power regulation and control. The power source, solar arrays, has been discussed above. The energy storage is in a battery. A regulated bus by means of a S4R (sequential switching & serial shunt regulator) is the best architecture; this is lightweight. Then, no BCR (Battery Charge Regulator) is needed (a heavy component). That leads to a light PCU (Power Conditioning Unit). The battery and its Battery Discharge Regulators are only occasionally needed. These components weigh, together, <5 kg. This is just for one ESA-*FIRI* telescope, without much equipment. Let us say that each of our *FIRI* satellites needs 7.5 kg.

The complete power system for one *FIRI* telescope will weigh 15 kg, producing 1500 W. Most components have TRL 9, but the solar panels have TRL 7.

TT&C, on-board data handling

The *Darwin* report (2000) mentions an antenna of 0.4 kg; an amplifier, receiver, modem of 10 kg; a computer of 5 kg and a harness of 3 kg. The total is 18.4 kg. By employing modern ASIC's and a lower-weight computer this could be brought back to 14 kg. For the telecommunications subsystem the ESA-*FIRI* report estimates 33 kg, at TRL level 9. In summary the total weight of the TT&C subsystem will be 14 kg and it will consume 36 Watt (*Darwin* 2000).

Data handling has been discussed in the payload section. Based on *Herschel* we deduce that when on-board data compression takes place, and no large data storage is needed, the weight of OBC plus mass memory will be of order 25 kg, at TRL level 9. Its power consumption may be 15 W (*Darwin* 2000).

Thermal sub-system

The functions to be provided consist of: (i) passive control by selection of surface properties, (ii) control of conduction paths and thermal capacities, and use of insulation, (iii) active control by heaters, louvers and shutters, refrigerators, (iv) absorptivity and emissivity of external surfaces should be controlled, (v) multi-layer Insulation (MLI) blankets may be used or single-layer radiation shields.

While L2 is a benign thermal environment, baffles and sunshields are necessary to reach the ~4 K needed for a direct-detection interferometer. The ESA-*FIRI* study assigned sunshield diameters 5× the telescope diameter. This would not be acceptable for the 3.5-m dishes of *FIRI*, so detailed studies are needed to pin down the numbers. This should take into account the number of layers, the temperature of each layer, the cryo-cooler and passive radiators to reach each temperature stage, etc. Since the shields will be too large to be easily stowed in the fairing, a mechanism to deploy the structures is needed. These mechanisms should be designed such that redundancy prevents failure.

The weightiest elements will be the sunshields and a cooling system for the detectors. ESA-*FIRI* (page 249) lists 173 kg of equipment for boom, hub, and the two telescopes. Relevant for one *FIRI* telescope are: a sunshield (29.1 kg), heaters (1.3 kg), WP-radiators (13.8 kg), MLI (2.5 kg). *FIRI*'s sunshield needs to be deployable and much larger than that for ESA-*FIRI*. By making use of the lightweight techniques described elsewhere in this section 60 kg may be sufficient.

The thermal sub-system of one *FIRI* satellite may weigh 75 kg, excluding cryo-cooling (see next section).

Payload complement

The heterodyne payload complement consists of local oscillator system, LO distribution system including central computer for the geometrical model, focal plane with detectors, common optics and IF amplifiers and the correlator system. We have shown in an earlier section that these are approximately 60 kg/satellite, i.e. 180 kg in total. Total power is more difficult to calculate but an estimate would be up to 200 W. Focal plane coolers will take another 300 W (current technology: the Japanese SMILES cooler) and adding 90 kg for cryostat, cooler and cooler electronics. These numbers will be brought down with development.

The direct detection system is comprised of identical light collectors and a beam combiner. The light collector payload consists of telescope optics, passive cooling elements, a cryocooler, and metrology system components. Based on the *SPECS* study, the light collector payload mass is 580 kg and its power requirement is 350 W. The beam combiner payload includes a single science instrument (Fourier Transform Spectrometer with multi-pixel detector arrays), a tether assembly, and metrology components. The beam combiner payload mass is 526 kg and its power requirement is 1250 W. All of the services provided by a traditional spacecraft bus, such as the communication system, are located at the beam combiner.

Metrology is common to all the proposed large missions, like *LISA, XEUS, Darwin* and *FIRI*. Of these the heterodyne interferometer on *FIRI* is the one with the least demanding knowledge requirements with respect to distance, followed by *XEUS* and the direct detection part of *FIRI*. However, even the heterodyne payload for *FIRI* needs (a posteriori) knowledge of the exact positions, so a full-blown metrology system is needed. *Darwin* (2000, p. 170) mentions 6.4 kg and 13 W/satellite for the metrology. EADS Astrium, TNO-TPD, SIOS INETI-LAER, and EADS CASA Espacio have performed "high precision optical metrology studies" capable of reaching 0.1 mm over 10s of metres.

The distributed correlator on board each satellite is estimated in the *ESPRIT* study as 5 kg/satellite.

The weight of the heterodyne instrument, including correlator and metrology sub-systems, is ~90 kg.

Weight/power of service module and instruments

Table 3 summarises the weight and power consumption in the heterodyne system.

**Table 3:** Weight and power summary.

| Subsystem | Weight (kg) | Power (W) |
|---|---|---|
| Att. Control | 65 | 50 |
| Tel., etc. | 100 | 0 |
| Power system | 15 | -1500 |
| TT&C; OBC | 40 | 50 |
| Thermal SS | 75 | 0 |
| Payload | 160 | 400 |
| Total | 455 | -1100 |

Undoubtedly quite a number of weight-contributing and power-consuming components have been neglected, but it is reassuring to see that a weight of 450 to 500 kg per satellite seems feasible. Concerning power requirements it is likely that providing 1500 W/satellite is amply sufficient, possibly less is needed.

A separate estimate of weight and power requirements for the beam combining satellite is needed. *Darwin*-2000 estimates for the "hub": 396 kg and 491 W (p. 171). This may be representative for *FIRI*'s hub as well.

**Table 4:** Direct detection system weight and power.

| Collector | | |
|---|---|---|
| | Mass | Power |
| Science Payload | 580 kg | 350 W |
| Mechanical/Structural | 395 kg | 0 W |
| Thermal | 143 kg | 50 W |
| Attitude Control System | 42 kg | 117 W |
| Command and Data Handling | 10 kg | 55 W |
| Communications | 10 kg | 17 W |
| Propulsion | 84 kg | 5000 W |
| Electrical Power System | 411 kg | 347 W |
| TOTAL | 1675 kg | 5936 W |

| Combiner | | |
|---|---|---|
| | Mass | Power |
| Science Payload | 526 kg | 1250 W |
| Mechanical/Structural | 469 kg | 0 W |
| Thermal | 214 kg | 50 W |
| Attitude Control System | 87 kg | 177 W |
| Command and Data Handling | 47 kg | 237 W |
| Communications | 103 kg | 122 W |
| Propulsion | 104 kg | 5000 W |
| Electrical Power System | 470 kg | 408 W |
| TOTAL | 2020 kg | 7244 W |

Table 4 summarises the weight and power in a direct detection interferometer according to the *SPECS* study.

Hall effect thrusters were baselined for use in *SPECS* with xenon as the propellant, accounting for the high propulsion power requirements. Two light-collecting telescopes and a beam combiner can be stacked and packaged in a canister for launch and deployment. With contingency and including the canister and "wet" mass components, the total observatory mass was estimated at 9560 kg (prior to any serious light-weighting studies and the introduction of cryo-coolers). In reality we expect the launch to be limited by volume, not mass.

Both the direct detection system and the heterodyne system are susceptible to EMC. *HIFI* – for comparison - was eventually very well shielded against EMC, after many tests, modeling and implementation of special measures. *FIRI* should not only be designed with mechanical and optical interfaces in mind, but from the start with EMC measures in place. This will highly simplify the testing and operation of the instrument. In the next years we will gain a lot more knowledge about the intensity of the solar wind and the cosmic ray flux in L2, and we can use the *Spitzer* experience. This will more precisely specify the amount of shielding necessary for the electronics and the direct detectors.

Specific environmental constraints

Normal measures will be taken to provide the right thermal environment for each *FIRI* component. See the subsection "Thermal subsystem". Here, the experience of building *Herschel* instruments seems adequate.

Current heritage (assumed bus) and TRL

Above we have given a description of many elements mounted in the service module. The *Herschel* SVM is too heavy to be considered for *FIRI*; fortunately, significant reductions can be achieved by use of micro-electronics, mechatronics and novel techniques. As far as possible, we have described the heritage and TRL, but a proper assessment phase is required to learn more about the applicability of new techniques.

Special requirements

The first major special requirement, common to every proposed large mission, is formation flying. *FIRI* would benefit from having data available from *GRACE, STEREO* and in the future *PROBA-3* as well as from ESA, NASA and industrial studies on this topic. The relaxed requirements with respect to *Darwin* may make it even possible to scale these studies to the *FIRI* case. However, the *FIRI* team does not have access to these data and therefore we consider formation flying as special requirement as well as a critical issue. Studies of formation flying should address the following:

- What accuracy is needed to keep control of the elements?
- What is the closest distance between any two elements; how is the risk of collisions treated?
- What is the best way to fill the UV plane in a reasonable amount of time?
- What propulsion system must be used to generate images with adequate UV plane filling?
- Can tethers save fuel? What are the risks?

The second major special requirement is the cooling of *FIRI*. This cooling consists of two components: for heterodyne observations, the mixers must be cooled down to 4 K. For direct detection, active cooling of the dishes and beam combiner allows for very sensitive direct detection measurements. While studies should show the exact requirements, we note that cooling is an issue for many proposed missions and that *FIRI* can benefit from existing studies. For *Darwin*, studies have been performed for vibration-less sorption micro-coolers. At first glance this seems very appropriate for the mixer cooling described above and the study results applied to *FIRI* could provide the detailed numbers on cooling power versus power consumption, mass and volume. Cooling the telescopes to 4 K is a difficult task, but the *SPECS* and *SPIRIT* studies show it to be feasible, and cooling a 3.5-m aperture to 4 K is expected to be demonstrated by *SPICA*. Confirmation of the numbers from these studies for *FIRI* is necessary.

6.6 Procurement approach and international partners

The final procurement and partnership agreements have to be negotiated between ESA, the instrument consortium and the partner agencies such as NASA and CSA. We propose the following: during the early assessment studies by industry and ESA, trade-offs for many aspects of *FIRI* will be done and it will lead to initial technical specifications for the interferometer as a whole and the elements themselves. These studies should not be limited to the straw-man design here but must be optimised to find the best solution to do the most science per Euro. In collaboration with the other space agencies, a plan will be drafted that will lead to coordinated open Invitations To Tender in ESA member states. ESA will assess the industrial proposals and choose the best ones. We assume that the spacecraft (including satellite bus) will be built by industry, with interface management done by ESA. The payload instruments may have several sub-systems that could be contracted out to space industry as well, but these sub-systems will be managed by the instrument consortium. Highly specialised design and construction heritage for instrumentation at these wavelengths exists only in instrument-building science institutes and consortia, which are usually nationally funded. Therefore careful thought needs to be given to how best to capture this knowledge if ESA procures instruments via industry. The procurement method needs study in Phase A. As an alternative, the R&D could be done with national funding, while ESA (and NASA/CSA) fund the instrument manufacture. ESA, CSA and NASA have a long tradition of collaboration with clear and distinct responsibilities for the agencies. This can be achieved by sub-dividing the spacecrafts and payloads in big chunks with easy interfaces. *LISA* and *JWST* are good examples of such divisions.

6.7 Critical issues

Although *FIRI* has not benefitted from years of industrial studies, other mission studies have encompassed much of what *FIRI* needs. There are, however, some open questions which need answers before *FIRI* enters Phase A:

- Direct detector array development to very low NEPs with high frequency read-out;
- Heterodyne detector development: can we approach the quantum limit whilst increasing IF bandwidth?
- LO development: how do we make reliable LOs at THz frequencies?
- Beam combiner: what is the design for a 4 K three/two-input beam combiner?
- Delay lines: how best to design optical delays employed at 4 K?
- Correlator: what is the best, most compact, power efficient cross-correlator?
- Telescopes: how do we make the best light-weight dish with off-axis movable secondary?
- Focal plane: how do we cool the mixers to 4 K?
- Telescopes: how do we cool dishes to 4 K?
- System: What is the optimal number and optimal temperature of light-collecting spacecraft?
- System: What is the spacecraft thermal design?
- System: What is the best propulsion, thruster, tether scheme for the whole system?
- System: What is the best metrology approach?
- System: What is the best optical design and how do we do stray-light suppression?
- System: What is the best AIV approach for a system that is extremely difficult to test on ground?
- System: How can we ensure an adequate number of guide stars?

In order to bring *FIRI* to the required level these studies are absolutely necessary. Science and technology institutes in Europe and North America are well equipped to tackle the fundamental issues like detector development. Space industry has the required capacity to solve the system issues. Detailed road maps can be produced easily in the assessment studies.

## 7. Science Operations and Archiving

ESA, its partners and the participating institutes will jointly set-up a ground segment taking care of all science operations. Data will be distributed electronically to the successful proposers and all data, i.e. all measured visibilities and housekeeping, will be archived by ESA and NASA. The proprietary time we anticipate is one year per observation, after which data are freely accessible to any interested astronomer. All data are available for calibration issues, however this data cannot be published without consent from the original proposer within 1 yr of acquisition. The calibration results can, however, be published.

We expect that time on *FIRI* would be given out as open time to the astrophysical community, with proposals evaluated through rigorous peer review led by ESA-appointed scientists.

## 8. Key Technology Areas

8.1 Payload TRL and technology development strategy

The TRL levels have been indicated as far as known. Generally they are satisfactorily high (usually above 6).

Needed technology development areas have been outlined at the end of Section 6. It is expected that for many of these questions the parallel developments taking place for missions like *Darwin, XEUS, LISA*, etc. will provide the answers. A few studies are already undertaken by the consortium partners, and others will be done in collaboration with the space agencies.

8.2 Mission and Spacecraft technology challenges

Section 6 describes the major technology challenges.

8.3 Mission schedule drivers

We expect that with all the development done for *Darwin* (including a technology demonstrator) the concept of free-flying will soon be mastered. Also configurations, $(u,v)$ filling and collision avoidance should be fully mastered. If not, technology development is needed here. High detector sensitivity is required for *FIRI*. We expect that ESA, NASA, and national funding agencies will make funding available for improving FIR detectors such that these can be used in any FIR mission, including *FIRI*. This concerns both coherent and incoherent detectors. Correlation and beam combination have, besides from the *ESPRIT, SPECS* and *SPIRIT* studies, been investigated for *ALMA* and for *Darwin*. The applicability of these technologies to *FIRI* deserves detailed study. Cooling, of the focal planes and/or of the complete payloads, poses a major problem to *FIRI*. Dedicated, in-depth studies including thorough thermal modelling have to be done in order to assess the limits of *FIRI*.

## 9. Preliminary Programmatics

9.1 Overall proposed mission management structure

ESA has a long tradition of cooperation with other space agencies, in particular with NASA and CSA. *FIRI* has always been envisaged as a collaborative mission (Ivison & Blain 2005), to be shared amongst several agencies. As an observatory-type mission, we propose that solutions similar to those implemented for *Herschel*, or for *HST* and *JWST*, should be investigated. We expect that ESA will be the body that oversees, and is responsible for, the interface management between instrument consortia and space industry. ESA would also set up a Science Working Group (SWG) to provide advice on all aspects of *FIRI* definition and development (mission studies, observatory specifications, choice of payload, Science Management Plan, etc.). The SWG should be set up from the beginning as an international body with all partners represented and a mix of instrumentation and science expertise – as independent and uncompromised as possible – chaired by an ESA-appointed *FIRI* Project Scientist.

9.2 Payload/Instrument Costs

Assumed share of payload costs to ESA

We assume that the scientific payload costs, will be met by national agencies within Europe, and NASA/CSA. ESA participation in the instrument consortia may be beneficial and could be discussed by the relevant parties.

Estimated non-ESA payload costs

The payload complement consists of two separate (heterodyne/direct detection) packages, each of which can be divided into several sub-systems. Building HIFI and the direct detection instruments for *Herschel* and *Planck* indicates that instrument costs are likely to around 100 M€ each (but considerable investment in relevant R&D will also be needed in the run-up to the mission). Here we assume a total of 300 M€ for the scientific instruments, including contingency.

9.3 Overall mission cost analysis

*FIRI* will be an L-class mission with a total cost to ESA of 640 M€, including contingency. Cooperation with NASA and CSA, who are suffering similarly tight budgetary constraints, can ensure *FIRI* remains within the L-class bracket. After a thorough Phase A study, the cost to ESA may be lowered.

**Table 5:** Mission cost breakdown (all figures in M€).

| Item | Meuro | ESA (60% of non-payload costs) | NASA (35% of all costs) | CSA (5% of all costs) | European Agencies (60% of payload instruments) |
|---|---|---|---|---|---|
| **Spacecraft and Agency Costs** | | | | | |
| Launcher | 125 | | | | |
| Pre-implementation | 5 | | | | |
| Ground Segment | 80 | | | | |
| ESA Project Costs | 50 | | | | |
| Spacecraft inc. telescopes | 560 | | | | |
| Contingency (30% on all of above) | 246 | | | | |
| **Total inc. contingency** | 1066 | 640 | 373 | 53 | |
| | | | | | |
| **Payload instruments** | 300 | | 105 | 15 | 180 |
| | | | | | |
| **Overall total** | 1366 | 640 | 478 | 68 | 180 |

Our cost estimate is summarised in Table 5. We will need an A5 ECA, with a cost of 125 M€. The Pre-Implementation Phase will cost around 5 M€, the Ground Segment 80 M€, and the ESA Internal Costs 50 M€. Spacecraft costs are scaled from the cost of the *SPICA* satellite/telescope (180 M€ per satellite, reduced by a factor of 1.3 to allow for commonality). This corresponds to 140 M€ per satellite, a total of 560 M€.

The proportion of costs to be borne by the participating agencies is something to be negotiated between the agencies. Here we assume the following: ESA supports 60% of all non-payload costs; NASA supports 35% of all costs; CSA supports 5% of all costs; European National agencies support 60% of the payload costs.

It should be borne in mind that that at this stage the costing methodology and division between participants is necessarily at an indicative level. One of the main objectives of the initial Assessment Study should be to produce more refined costings for all aspects of the mission, and to investigate division of activities between the participants that will keep the technical and managerial interfaces as simple as possible. The current exercise illustrates that the mission is feasible within the ESA L-class envelope with the suggested level of international participation.

## 10. Communications and Outreach

With the general public

As an high-resolution imaging mission, *FIRI* will generate the kind of headline-grabbing images produced by, for example, the *HST* or the *Huygens* lander – colourful high-fidelity images, with potentially collosal impact. Two of the key science cases, the formation of extra-solar Earth-like planets and the hunt for merging super-massive black holes, have visceral appeal to the general public and these offer opportunities to revisit *FIRI* achievements on a regular basis as new discoveries are made. From a technological perspective the concept of several satellites, each the size of a small bus, performing an orbital ballet well beyond the orbit of the Moon is both easily understood and obviously impressive to the lay public and will resonate with a generation who may be seeing the first astronauts walking on the Moon for almost half a century.

Although complicated in detail, the basic techniques and raw angular resolution of *FIRI* are easy to illustrate graphically. The project will develop suitable material for distribution to the media and maintain a publicly accessible portal via the Web for distribution and download of suitable materials. Special emphasis will be given to offering a dialogue with the interested public, rather than just feeding them a diet of "facts".

With educators

Electro-magnetic theory is a crucial element in physics education and interference (e.g. Newton's rings) are often used in high school to illustrate the wave theory of light. The project will work with educators to develop material (classroom exercises, low-cost experiments and demonstrations) which can be used to show *FIRI* as a dramatic example of how theoretical physics and practical engineering can be applied in the real world.

With the scientific community

For the next generation of young astronomers, the value of interferometric techniques will already be well established, a heritage built on facility instruments such as ALMA and VLTI. To take advantage of this heritage the project will hold workshops, training programmes and seminars, etc., to ensure that the full potential of the mission is understood by a wide population of users. This will assure proposals of very high quality which will fully exploit this unique resource.

## *Participating Institutes*

Austria

University of Innsbruck

France

CEA, Saclay
CESR, Toulouse
IRAM, Grenoble
Universite Bordeaux 1
LERMA, Paris

Germany

Argelander Institute, University of Bonn
MPE, Garching
MPIA, Heidelberg
MPIfR, Bonn

Italy

INAF, Astronomical Observatory of Rome
CISAS, University of Padova
INAF, Astronomical Observatory of Padova
IFSI-INAF, Rome

Netherlands

Rijksuniversiteit Groningen
SRON, Groningen
University of Leiden

Spain

Instituto de Astrofisica de Canarias, Tenerife
Instituto de Estructura de la Materia, DAMIR. CSIC, Madrid
Observatorio Astronomico Nacional, Alcala de Henares

Sweden

Onsala Space Observatory
Stockholm University

Switzerland

Swiss Federal Institute of Technology (ETH), Zurich

UK

Durham University
Imperial College London
Open University
Rutherford Appleton Laboratory (STFC)
UK Astronomy Technology Centre (STFC), Edinburgh
University of Cardiff
University of Edinburgh
University of Hertfordshire
University College London (including MSSL, Dorking)
University of Manchester
University of Oxford
University of St Andrews
University of Sussex

Canada

McGill University
McMaster University
National Research Council, HIA
University of Calgary
Université Laval
Universit of Lethbridge
Université de Montréal
University of Waterloo
University of Western Ontario

United States of America

Representatives of the following US institutes have participated in writing this proposal, but many other US institutes have been engaged in FIR mission planning, and in the *SPIRIT* and *SPECS* mission studies in particular. They will have an opportunity to participate in the *FIRI* Assessment Phase should NASA choose to support the *FIRI* study.

California Institute of Technology
Cornell University
Jet Propulsion Laboratory, Caltech
Johns Hopkins University
NASA Goddard Space Flight Center
Naval Research Laboratory
Ohio State University
University of Massachusetts
University of Michigan

*Supporting statements*

Formal statements from NASA and CSA, excised for this astro-ph posting.